\documentclass{article}

\usepackage{arxiv}

\usepackage[utf8]{inputenc} % allow utf-8 input
\usepackage[T1]{fontenc}    % use 8-bit T1 fonts
\usepackage{hyperref}       % hyperlinks
\usepackage{url}            % simple URL typesetting
\usepackage{booktabs}       % professional-quality tables
\usepackage{amsfonts}       % blackboard math symbols
\usepackage{nicefrac}       % compact symbols for 1/2, etc.
\usepackage{microtype}      % microtypography
\usepackage{lipsum}		% Can be removed after putting your text content
\usepackage{graphicx}
\usepackage[square,numbers]{natbib}
\usepackage{doi}

\usepackage{lipsum}		% Can be removed after putting your text content
\usepackage{amsmath}
\usepackage{booktabs}
\usepackage{xcolor}
\usepackage{colortbl}
\usepackage[nolist]{acronym}
\usepackage{listings} % source code listings
\usepackage[T1]{fontenc}
\usepackage{hyperref}
\usepackage{enumitem}
\usepackage[ruled, linesnumbered]{algorithm2e}
\usepackage{graphicx}
\usepackage{textcomp}
\usepackage{hyperref}
\usepackage{subcaption}
\usepackage{pdfpages}
\usepackage{forest}
\usepackage{xcolor}
\usetikzlibrary{angles}
\definecolor{drawColor}{RGB}{128 128 128}
\newcommand{\circleSize}{0.25em}
\newcommand{\angleSize}{0.8em}
\forestset{
	/tikz/mandatory/.style={
		circle,fill=drawColor,
		draw=drawColor,
		inner sep=\circleSize
	},
	/tikz/optional/.style={
		circle,
		fill=white,
		draw=drawColor,
		inner sep=\circleSize
	},
	featureDiagram/.style={
		for tree={
			text depth = 0,
			parent anchor = south,
			child anchor = north,
			draw = drawColor,
			edge = {draw=drawColor},
		}
	},
	/tikz/abstract/.style={
		fill = blue!85!cyan!5,
		draw = drawColor
	},
	/tikz/concrete/.style={
		fill = blue!85!cyan!20,
		draw = drawColor
	},
	mandatory/.style={
		edge label={node [mandatory] {} }
	},
	optional/.style={
		edge label={node [optional] {} }
	},
	or/.style={
		tikz+={
			\path (.parent) coordinate (A) -- (!u.children) coordinate (B) -- (!ul.parent) coordinate (C) pic[fill=drawColor, angle radius=\angleSize]{angle};
		}	
	},
	/tikz/or/.style={
	},
	alternative/.style={
		tikz+={
			\path (.parent) coordinate (A) -- (!u.children) coordinate (B) -- (!ul.parent) coordinate (C) pic[draw=drawColor, angle radius=\angleSize]{angle};
		}	
	},
	/tikz/alternative/.style={
	},
	/tikz/placeholder/.style={
	},
	collapsed/.style={
		rounded corners,
		no edge,
		for tree={
			fill opacity=0,
			draw opacity=0,
			l = 0em,
		}
	},
	/tikz/hiddenNodes/.style={
		midway,
		rounded corners,
		draw=drawColor,
		fill=white,
		minimum size = 1.2em,
		minimum width = 0.8em,
		scale=0.9
	},
}
\title{MulTi-Wise Sampling: Trading Uniform T-Wise Feature Interaction Coverage for Smaller Samples}

\author{ \href{https://orcid.org/0000-0001-7652-6525}
		{\includegraphics[scale=0.06]{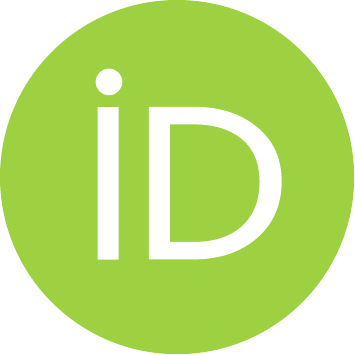}\hspace{1mm}
		Tobias~Pett} \\ %\thanks{Use footnote for providing further
		Karlsruhe Institutte of Technology\\
		Karlsruhe, Germany \\
		\texttt{tobias.pett@kit.edu} \\
	\And
		\href{https://orcid.org/0000-0001-7077-7091}
		{\includegraphics[scale=0.06]{orcid.pdf}\hspace{1mm}
		Sebastian~Krieter} \\
		Paderborn University \\
		Germany \\
		\texttt{sebastian.krieter@uni-paderborn.de} \\
	\And
		\href{https://orcid.org/0000-0001-8069-9584}
		{\includegraphics[scale=0.06]{orcid.pdf}\hspace{1mm}
		Thomas~Thüm} \\
		Paderborn University \\
		Germany \\
		\texttt{thomas.thüm@uni-paderborn.de} \\
	\And
		\href{https://orcid.org/0000-0002-7153-761X}
		{\includegraphics[scale=0.06]{orcid.pdf}\hspace{1mm}
		Ina~Schaefer} \\
		Karlsruhe Institutte of Technology\\
		Karlsruhe, Germany \\
		\texttt{ina.schaefer@kit.edu} \\
}

\renewcommand{\headeright}{MulTi-Wise Sampling: Trading Uniform T-Wise Feature Interaction Coverage for Smaller Samples}
\renewcommand{\undertitle}{This is an extension for the short paper: \\
 							\emph{MulTi-Wise Sampling: Trading Uniform T-Wise Feature Interaction Coverage for Smaller Samples} \\
							Published at 28th acm International Systems and Software Product Line Conference under the doi:\\
							\emph{10.1145/3646548.3672589}}
\renewcommand{\shorttitle}{}

\hypersetup{
pdftitle={MulTi-Wise Sampling: Trading Uniform T-Wise Feature Interaction Coverage for Smaller Samples},
pdfsubject={Computer Science - Software Engineering},
pdfauthor={Tobias~Pett, Sebastian~Krieter, Thomas~Thüm, Ina~Schaefer},
pdfkeywords={t-wise coverage, software-product lines, spl testing, sampling},
}

\begin{document}

%% Import utility packages part 2
%% tex file used to define acronyms

% Color Scheme http://colorschemedesigner.com/#3w40I--ALK-K-
% Base Color of the OVGU INF logo, tetraed, -45
%% KIT Green
\definecolor{kitGreen1}{HTML}{00876c}
\definecolor{kitGreen2}{HTML}{b7fff5}
\definecolor{kitGreen3}{HTML}{6fffec}
\definecolor{kitGreen4}{HTML}{27ffe2}
\definecolor{kitGreen5}{HTML}{007162}
\definecolor{kitGreen6}{HTML}{004b41}
%% =========================================
%% KIT Blue
\definecolor{kitBlue1}{HTML}{4664aa}
\definecolor{kitBlue2}{HTML}{d9dfef}
\definecolor{kitBlue3}{HTML}{b3c0df}
\definecolor{kitBlue4}{HTML}{8ca1d0}
\definecolor{kitBlue5}{HTML}{354b7f}
\definecolor{kitBlue6}{HTML}{233255}
%% =========================================
%% KIT Black
\definecolor{kitBlack1}{HTML}{404040}
\definecolor{kitBlack2}{HTML}{000000}
\definecolor{kitBlack3}{HTML}{7f7f7f}
\definecolor{kitBlack4}{HTML}{595959}
\definecolor{kitBlack5}{HTML}{262626}
\definecolor{kitBlack6}{HTML}{0d0d0d}
%% =========================================
%% KIT May Green
\definecolor{kitMayGreen1}{HTML}{77a200}
\definecolor{kitMayGreen2}{HTML}{e8f2d7}
\definecolor{kitMayGreen3}{HTML}{d2e4ae}
\definecolor{kitMayGreen4}{HTML}{bbd786}
\definecolor{kitMayGreen5}{HTML}{69882d}
\definecolor{kitMayGreen6}{HTML}{465b1e}
%% =========================================
%% KIT Purple
\definecolor{kitPurple1}{HTML}{a3107c}
\definecolor{kitPurple2}{HTML}{f9c3eb}
\definecolor{kitPurple3}{HTML}{f386d6}
\definecolor{kitPurple4}{HTML}{ed4ac2}
\definecolor{kitPurple5}{HTML}{7a0c5d}
\definecolor{kitPurple6}{HTML}{52083e}
%% =========================================
%% KIT Brown
\definecolor{kitBrown1}{HTML}{a7822e}
\definecolor{kitBrown2}{HTML}{df9b1b}
\definecolor{kitBrown3}{HTML}{f9ebd1}
\definecolor{kitBrown4}{HTML}{f4d7a2}
\definecolor{kitBrown5}{HTML}{eec474}
\definecolor{kitBrown6}{HTML}{6f4e0e}
%% =========================================
%% KIT Cyan
\definecolor{kitCyan1}{HTML}{079ede}
\definecolor{kitCyan2}{HTML}{d3ecf9}
\definecolor{kitCyan3}{HTML}{a7d9f3}
\definecolor{kitCyan4}{HTML}{7bc7ec}
\definecolor{kitCyan5}{HTML}{187aaa}
\definecolor{kitCyan6}{HTML}{105172}
%% =========================================
%% KIT Gray
\definecolor{kitGray1}{HTML}{d9d9d9}
\definecolor{kitGray2}{HTML}{c3c3c3}
\definecolor{kitGray3}{HTML}{a3a3a3}
\definecolor{kitGray4}{HTML}{6d6d6d}
\definecolor{kitGray5}{HTML}{363636}
\definecolor{kitGray6}{HTML}{161616}
%% =========================================
%% KIT Yellow
\definecolor{kitYellow1}{HTML}{fce500}
\definecolor{kitYellow2}{HTML}{fffacb}
\definecolor{kitYellow3}{HTML}{fff698}
\definecolor{kitYellow4}{HTML}{fff164}
\definecolor{kitYellow5}{HTML}{bdac00}
\definecolor{kitYellow6}{HTML}{7e7200}
%% =========================================
%% KIT White
\definecolor{kitWhite1}{HTML}{ffffff}
\definecolor{kitWhite2}{HTML}{f2f2f2}
\definecolor{kitWhite3}{HTML}{d9d9d9}
\definecolor{kitWhite4}{HTML}{bfbfbf}
\definecolor{kitWhite5}{HTML}{a6a6a6}
\definecolor{kitWhite6}{HTML}{7f7f7f}

\definecolor{background}{named}{white}
\definecolor{bgborder}{named}{black}
\definecolor{comment}{named}{red}
\definecolor{hint}{named}{kitBlue1}

\definecolor{blue}{named}{kitBlue1}
\definecolor{green}{named}{kitGreen1}
\definecolor{red}{named}{red}
\definecolor{orange}{named}{kitBrown1}

\definecolor{pdflinkcolor}{named}{kitBlue3}
\definecolor{pdfcitecolor}{named}{kitBlue3}

%% check for documentclass option
\newcommand{\checkAuthormode}[1]{\ifnum\myAuthormode=1{#1}\fi}
%% commands for todos in various forms
\newcommand{\todoUC}[1]{{\color{comment}\textit{[#1]}}}
\newcommand{\todoTPUC}[1]{{\color{kitBlue1}\textit{[\textbf{@Tobi P.:} #1] \newline}}}
\newcommand{\todoSKUC}[1]{{\color{kitBlue2}\textit{[\textbf{@Krieter:} #1] \newline}}}
\newcommand{\todoISUC}[1]{{\color{kitBlue3}\textit{[\textbf{@Ina:} #1] \newline}}}
\newcommand{\todoTTUC}[1]{{\color{kitBlue4}\textit{[\textbf{@Thomas:} #1] \newline}}}
\newcommand{\todoTHUC}[1]{{\color{kitBlue5}\textit{[\textbf{@Tobias H.:} #1] \newline}}}
\newcommand{\todoAllUC}[1]{{\color{kitBrown3}\textit{[\textbf{@ALL:} #1] \newline}}}
\newcommand{\todoCiteUC}[1]{{\color{comment} ~Cite #1}}
%% indicate that text will follow here
\newcommand{\todotsUC}{\todo{\ldots}}
%% remarks
\newcommand{\REMARKUC}[1]{{\color{kitBlue1} #1 \newline}}
\newcommand{\EXISTING}[1]{{\color{kitBrown3} \textit{#1}}}
%% command to indicate new text sections
\newcommand{\new}[1]{{\color{kitMayGreen1}#1}}

%% todo commands using the check author macro to be displayed only in the author version of the document
\newcommand{\todo}[1]{\checkAuthormode{\todoUC{#1}}}
\newcommand{\todoTP}[1]{{\checkAuthormode{\todoTPUC{#1}}}}
\newcommand{\todoSK}[1]{{\checkAuthormode{\todoSKUC{#1}}}}
\newcommand{\todoIS}[1]{{\checkAuthormode{\todoISUC{#1}}}}
\newcommand{\todoTT}[1]{{\checkAuthormode{\todoTTUC{#1}}}}
\newcommand{\todoTH}[1]{{\checkAuthormode{\todoTHUC{#1}}}}
\newcommand{\todoAll}[1]{{\checkAuthormode{\todoAllUC{#1}}}}
\newcommand{\todoCite}[1]{{\checkAuthormode{\todoCiteUC{#1}}}}
%% indicate that text will follow here
\newcommand{\todots}{\checkAuthormode{\todotsUC}}
%% remarks
\newcommand{\REMARK}[1]{{\checkAuthormode{\REMARKUC{#1}}}}
%% reworked part
\newcommand{\rework}[1]{{\color{kitGreen1} #1}}

%% autorefs
\newcommand{\autorefSec}[1]{Section #1}
%Example Environment
%\newcounter{EXAMPLE}[section]
%\newenvironment{EXAMPLE}
%% begin env
%{\refstepcounter{EXAMPLE}\par\medskip
%	\noindent \textbf{\textit{Example~\thesection.\theEXAMPLE}} 
%	\begin{itshape}	
%}
%%end Env
%{\end{itshape}
%}
%\newcommand{\EXAMRef}[1]{\color{black} Example~\ref{#1}\color{black}}

%% environment to mark changed parts of the paper
\newenvironment{ChangedEnv}{
	\color{green3}
}
{\color{black} \vspace{0.3cm}
}
%\renewcommand\lstlistingname{Quelltext}
%% example code style using java 
%\lstdefinestyle{java}{
%code formatting
%	language=Java,
%	tabsize=4,
%	breaklines=false,
%	basicstyle=\fontfamily{pcr}\footnotesize\selectfont,
%	commentstyle=\fontshape{it}\color{darkgray}\selectfont,
%	keywordstyle=\fontseries{b}\selectfont,
%	stringstyle=\fontfamily{cmr}\selectfont,
%%line numbering
%	numbers=left,
%	numberstyle=\footnotesize,
%%frame properties
%	captionpos=b,
%	frame=single,%trblTRBL
%	framesep=3pt,
%	xleftmargin=4pt,
%	xrightmargin=4pt,
%	rulecolor=\color{bgborder},
%}
%% another example to visusalize feature models as cnf
%\lstdefinestyle{FM}{
%	backgroundcolor=\color{lightgray!30},   
%	rulecolor=\color{lightgray}, 
%	frame=single,
%	commentstyle=\color{green1},
%	keywordstyle=\color{red1},
%	numberstyle=\tiny\color{black},
%	stringstyle=\color{black},
%	basicstyle=\ttfamily\footnotesize,
%	breakatwhitespace=false,         
%	breaklines=true,                 
%	captionpos=t,                    
%	keepspaces=true,                 
%	%numbers=left,                    
%	numbersep=5pt,                  
%	showspaces=false,                
%	showstringspaces=false,
%	showtabs=false,                  
%	tabsize=1,
%	literate=
%	{-}{$\neg$}{1}
%	{&}{,}{1}
%	{|}{$\lor$}{1},
%	mathescape = true,
%	morekeywords={include}
%}

% propositional formulas
\newcommand{\pand}{\wedge}
\newcommand{\por}{\vee}
\newcommand{\pnot}{\neg}
\newcommand{\pequals}{\Leftrightarrow}
\newcommand{\pimplies}{\Rightarrow}
\newcommand{\pnimplies}{\nRightarrow}
\newcommand{\patmostone}{\mbox{\textit{atmost1}}}
\newcommand{\pchooseone}{\mbox{\textit{choose1}}}
\newcommand{\abs}[1]{|#1|}
\newcommand{\bigPipe}{~\big|~}
\newcommand{\setOpen}{~\big\{~}
\newcommand{\setClose}{~\big\}}
\newcommand{\listOpen}{~\big[~}
\newcommand{\listClose}{~\big]}
\newcommand{\tupleOpen}{~\big(~}
\newcommand{\tupleClose}{~\big)}

% variable allotment
\newcommand{\T}{T{ }}
% mathematical definitions and theorems
\newtheorem{DEFINITION}{Definition}[section]
\newtheorem{THEOREM}{Theorem}[section]
\newtheorem{LEMMA}{Lemma}[section]
%% this is the place to define commands for important terms of the paper
%\newcommand{\car}{\texttt{Car}}
%\newcommand{\carbody}{\texttt{Carbody}}
%\newcommand{\radio}{\texttt{Radio}}
%\newcommand{\gearbox}{\texttt{Gearbox}}
%\newcommand{\ports}{\texttt{Ports}}
%\newcommand{\navigation}{\texttt{Navigation}}
%\newcommand{\bluetooth}{\texttt{Bluetooth}}
%\newcommand{\manual}{\texttt{Manual}}
%\newcommand{\automatic}{\texttt{Automatic}}
%\newcommand{\usb}{\texttt{USB}}
%\newcommand{\cd}{\texttt{CD}}
%-----------------------------------------------------
%% Definition of constant terms 
\newcommand*{\mulTiWise}[1]{MulTi-Wise Sampling}
\newcommand*{\yasa}[1]{\textsc{yasa#1}}
\newcommand*{\busybox}[1]{BusyBox}
\newcommand*{\fiasco}[1]{Fiasco}
\newcommand*{\uclibc}[1]{uCLibc-ng}
\newcommand{\soletta}[1]{Soletta}

%%%%%% Eval 
\newcommand*{\setupOne}[1]{\emph{Exp1}}
\newcommand*{\setupTwo}[1]{\emph{Exp2}}
\newcommand*{\setupThree}[1]{\emph{Exp3}}
\newcommand*{\setupFour}[1]{\emph{Exp4}}
\newcommand*{\setupFive}[1]{\emph{Exp5}}
\newcommand*{\setupSix}[1]{\emph{Exp6}}
\newcommand*{\setupSev}[1]{\emph{Exp7}}

%% Definition of mathmatical terms --> usage in text requires math environment
%% feature model
\newcommand{\model}{\mathcal{M}}

%% Feature space
\newcommand{\featureSpaceLHS}{\mathcal{F}}
\newcommand{\featureSpaceRHS}{\{f_0, f_1, \dots, f_i\}}
\newcommand{\featureI}{f_i}
\newcommand{\feature}{f}
\newcommand{\featureSpaceCount}{o}

%% dependencie space
\newcommand{\depSpaceLHS}{\mathcal{D}}
\newcommand{\depSpaceRHS}{\{d_0, d_1, \dots, d_i\}}
\newcommand{\depSpaceI}{d_i}
\newcommand{\depSpaceCount}{p}

%% Literals of model
\newcommand{\litSpaceLHS}{\mathcal{L}(\model)}
\newcommand{\litSpace}{\mathcal{L}}
\newcommand{\literal}{f}
\newcommand{\litI}{\literal_i}
\newcommand{\litN}{n}
\newcommand{\litSpaceRHS}{\setOpen \literal_{0}, \literal_{1}, \dots, \literal_{n} \bigPipe \litN \in \abs{\featureSpaceLHS} \setClose}
%%
%% t-wise interaction size (t=2, t=3 usw.)
\newcommand{\twise}{t}
\newcommand{\twiseText}{t~}
%% t-wise feature tuple
\newcommand{\tTuple}{tTuple}
%%
%% feature permutation
\newcommand{\featurePermutation}{2^{\featureSpaceLHS}}
%%
%% Configuration space
\newcommand{\confSpace}{\mathcal{C}}
%% configuration 
\newcommand{\conf}{C}
%% selected feature set
\newcommand{\fsel}{\featureSpaceLHS_{sel}}
%% deselected feature set
\newcommand{\fdes}{\featureSpaceLHS_{des}}
%% undecided feature set
\newcommand{\fund}{\featureSpaceLHS_{und}}
%%
%% sample 
\newcommand{\samp}{S}
%% 
%% t-wise interactions of the model LHS
%\newcommand{\intersetLHS}{\mathcal{I}(\twise,\model, \featureSpaceLHS)}
\newcommand{\intersetLHS}[3]{\mathcal{I}(#1,#2,#3)}
\newcommand{\intersetLHSNoParam}{\mathcal{I}}
%% interaction set of the model RHS
%\newcommand{\intersetRHS}{\setOpen i \in 2^{\litSpaceLHS} \bigPipe \abs{i} = \twise \pand \exists \conf \in \confSpace(\model) : i \in \conf)\setClose}
\newcommand{\intersetRHS}{\setOpen (\feature_{1}, \dots \feature_{t}) \bigPipe (\feature_{1}, \dots \feature_{t}) \in \conf(\model) \setClose}
%%
%% single t-interaction
\newcommand{\interactionLHS}{I}
\newcommand{\interactionRHS}{\tupleOpen \feature_{1}, \dots, \feature_{t} \tupleClose}
%%
%% t-wise interactions by the sample LHS
\newcommand{\interInSamLHS}{\mathcal{I}(\twise, \model, \featureSpaceLHS, \samp)}
%% t-wise interactions by the sample RHS
\newcommand{\interInSamRHS}{\setOpen (\feature_{1}, \dots \feature_{t}) \bigPipe~\text{there exists}~\conf \in \samp~\text{such as}~(\feature_{1}, \dots \feature_{t}) \subseteq \conf \setClose}
%%
%\newcommand{\interInSamSetLHS}{\mathcal{I}(\twise, \model, \sampleWindowLHS)}
%\newcommand{\interInSamSetRHS}{\listOpen \literal_{1}, \dots \literal_{t} \listClose \allowbreak ~\text{there exists}~\conf \in \sampleUnion~\text{such as}~\setOpen \literal_{1}, \dots \literal_{t} \setClose \subseteq \conf \setClose}
%%
%% one-shot coverage LHS
\newcommand{\osCoverLHS}{\text{coverage}~(\twise, \model, \samp)}
%% one-shot coverage RHS
\newcommand{\osCoverRHS}{\frac{\abs{\interInSamLHS}}{\abs{\intersetLHS{\twise}{\model}{\featureSpaceLHS}}}}
%%
%% t-wise feature interaction group Left Hand Side
\newcommand{\tGroupLHS}{TG}
%% t-wise feature interaction group Right Hand Side
\newcommand{\tGroupRHS}{\setOpen \feature_1, \feature_2, \dots, \feature_n \setClose}
%% set of all t-wise feature interaction groups left hand side
\newcommand{\tGroupSetLHS}{\mathcal{TG}}
%% set of all t-wise feature interaction groups right hand side
\newcommand{\tGroupSetRHS}{\setOpen \tGroupLHS_{1}, \dots, \tGroupLHS_{n}, \tGroupDefault \setClose}
%% t-wise feature interaction group Default
\newcommand{\tGroupDefault}{TGD}
%-------------------------------------------------------------------
%% static variables
%% Subject Systems
\newcommand{\listSubjectsystems}{\busybox{}, \fiasco{}, \soletta{}, and \uclibc{}}
\newcommand{\listSubjectsystemsFoot}{\busybox{}: \busyboxdata, \fiasco{}: \fiascodata, \soletta{}:\solettadata, \uclibc{}: \uclibcdata}
%------------------------------------------------------------------
%%%%%% Acks 
%\grantsponsor{⟨sponsorID⟩}{⟨name⟩}{⟨url⟩}
%\grantnum[⟨url⟩]{⟨sponsorID⟩}{⟨number⟩}

%%%% URLs for grants
%% Helmholz Association
%\newcommand{\grantHgf}{\grantsponsor{hgf}{Helmholtz Association (HGF)}{https://www.helmholtz.de/en/}}
%\newcommand{\grandNumHgf}{\grantnum{sofDCar}{19S21002}}

%% KASTEL Security Research Labs
\newcommand{\grantKASTEL}{\grantsponsor{kastel}{Engineering Secure Systems of the Helmholtz Association (HGF) and by KASTEL Security Research Labs}{\hgfUrl}}
\newcommand{\grantNumKASTEL}{\grantnum{kastel}{46.23.03}}

%% Ministry Science, Research, and Arts Ba-Wü (ICM)
\newcommand{\grantICM}{\grantsponsor{icm}{Ministry of Science, Research and Arts of the Federal State of Baden-Württemberg}{\icmUrl}}

%% Federal Ministry for Economic Affairs and Climate Action (SofDCar)
\newcommand{\grantSofDCar}{\grantsponsor{sofDCar}{SofDCar, which is funded by the German Federal Ministry for Economic Affairs and Climate Action}{\sdcUrl}}
\newcommand{\grantNumSofDCar}{\grantnum{sofDCar}{19S21002}}

%% Example URL Def.
%\urldef{\exampleURL}\url{https://example-urls.de}

\urldef{\busyboxdata}\url{https://github.com/TUBS-ISF/busybox-case_study}
\urldef{\solettadata}\url{https://github.com/TUBS-ISF/soletta-case-study}
\urldef{\toyboxdata}\url{https://github.com/TUBS-ISF/toybox-case-study}
\urldef{\uclibcdata}\url{https://github.com/TUBS-ISF/uclibc-case-study}
\urldef{\fiascodata}\url{https://github.com/TUBS-ISF/fiasco-case-study}

\urldef{\busyboxLink}\url{https://www.busybox.net/}
\urldef{\solettaLink}\url{https://github.com/solettaproject/soletta}
\urldef{\toyboxLink}\url{https://github.com/landley/toybox}
\urldef{\uclibcLink}\url{https://github.com/wbx-github/uclibc-ng/}
\urldef{\fiascoLink}\url{https://github.com/kernkonzept/fiasco}

\urldef{\fideLink}\url{https://featureide.github.io/}
\urldef{\carFmLink}\url{https://github.com/FeatureIDE/FeatureIDE/tree/develop/plugins/de.ovgu.featureide.examples/featureide_examples/FeatureModels/Car}

%% where is linux used
\urldef{\linuxUse}\url{https://www.tecmint.com/big-companies-and-devices-running-on-gnulinux/}

%% Linux Foundation publications
\urldef{\linuxPubs}\url{https://www.linuxfoundation.org/resources/publications/}

%% Linux foundation linux history report
\urldef{\linuxHistoryRep}\url{https://www.linuxfoundation.org/wp-content/uploads/2020_kernel_history_report_082720.pdf}

%% Linux Foundation anual report 2021
\urldef{\linuxAnualRep}\url{https://linuxfoundation.org/wp-content/uploads/2021_LF_Annual_Report_010222.pdf}

%% reference to testing of linux
\urldef{\testingLinuxA}\url{https://www.linuxjournal.com/article/7445}
\urldef{\testingLinuxB}\url{https://people.netfilter.org/hawk/presentations/ifdef2016/ifdef_FOSD2016.pdf}

%% replication packages
%\urldef{\implementationLink}\url{Anonymous}
\urldef{\implementationLink}\url{https://doi.org/10.5281/zenodo.11654696}

\urldef{\dataLink}\url{https://doi.org/10.5281/zenodo.11082621}

\urldef{\digitalOcean}\url{https://www.digitalocean.com/community/tutorials/how-to-install-java-with-apt-on-ubuntu-20-04}
\urldef{\microSoft}\url{https://learn.microsoft.com/en-us/windows/wsl/install}

\urldef{\addSubjectSystem}\url{https://github.com/SoftVarE-Group/feature-model-benchmark}

%%%% URLs for grants
%% Helmholz Association
\urldef{\hgfUrl}\url{https://www.helmholtz.de/en/}
%% KASTEL Security Research Labs

%% Ministry Science, Research, and Arts Ba-Wü
\urldef{\sdcUrl}\url{https://www.bundesregierung.de/breg-en/federal-government/ministries/ministry-for-economic-affairs-and-climate-action}

%% Federal Ministry for Economic Affairs and Climate Action
\urldef{\icmUrl}\url{https://mwk.baden-wuerttemberg.de/en/the-ministry}

%% Feature Model Benchmark by Sundermann on GitHub
\urldef{\fmbUrl}\url{https://github.com/SoftVarE-Group/feature-model-benchmark}

%% FeatureIDE 
\urldef{\fideUrl}\url{https://featureide.github.io/}

%% Contains the content for the footnotes of this paper

\newcommand*{\footReplication}{\implementationLink}
\newcommand*{\footResults}{\dataLink}
\newcommand*{\footFIDE}{\fideUrl}
\newcommand*{\footAddSubjectSystem}{\addSubjectSystem}
\newcommand*{\footArxiv}{\textcolor{kitPurple5}{Link to Arxiv Paper}}

\maketitle

\begin{abstract}
	Ensuring the functional safety of highly configurable systems often requires testing representative subsets of all possible configurations to reduce testing effort and save resources.
The ratio of covered t-wise feature interactions (i.e., T-Wise Feature Interaction Coverage) is a common criterion for determining whether a subset of configurations is representative and capable of finding faults.
Existing t-wise sampling algorithms uniformly cover t-wise feature interactions for all features, resulting in lengthy execution times and large sample sizes, particularly when large t-wise feature interactions are considered (i.e., high values of t). 
In this paper, we introduce a novel approach to t-wise feature interaction sampling, questioning the necessity of uniform coverage across all t-wise feature interactions, called \emph{\mulTiWise{}}.
Our approach prioritizes between subsets of critical and non-critical features, considering higher t-values for subsets of critical features when generating a t-wise feature interaction sample. 
We evaluate our approach using subject systems from real-world applications, including \busybox{}, \soletta{}, \fiasco{}, and \uclibc{}.
Our results show that sacrificing uniform t-wise feature interaction coverage between all features reduces the time needed to generate a sample and the resulting sample size.
Hence, \mulTiWise{} Sampling offers an alternative to existing approaches if knowledge about feature criticality is available.
\end{abstract}

% keywords can be removed

%
% Introduction
\section{Introduction}
%Modern cyber-physical systems, such as cars, trains, ships, and airplanes, play a pivotal role in our daily lives, transporting passengers at high relative velocities.
%These systems are characterised by their critical safety requirements, high configurability, high system complexity, and frequent update cycles. 
Nowadays configurable systems are highly complex, evolve frequently, and appear in safety-critical areas such as passenger transportation, leading to strict requirements for the functional safety of those systems. 
Automotive systems are prime examples of highly configurable systems for which functional safety must be assured throughout their life cycle.
%System testing is important to assure the functional safety of highly configurable systems~\cite{AmO+:16, M10}.
To assure the functional safety of highly configurable systems, system testing is important~\cite{AmO+:16, M10}.
However, system testing often mandates a trade-off between efficient test execution (i.e., the time it takes to execute all test cases) and test coverage (i.e., how many system configurations were covered by the testing procedure)~\cite{DMG:07, FF:06, Mey+:IEEESoftware14, PBL05, PTR+:SPLC19}.
This trade-off is even more severe for configurable systems since test cases must be executed on multiple system configurations. 
A system configuration is a selection of configuration options (i.e., features) of the configurable system~\cite{ABKS13}.  
Thorough testing would execute all test cases on all possible system configurations to achieve the highest possible system coverage. 
However, executing all test cases on all possible system configurations is not feasible in practice because of the combinatorial explosion problem~\cite{ER:IST11, LKL:SPLC12}. 
For instance, the analysis of JPHipster\cite{HNA+:EMSE19} has shown that executing all test cases for a system with 48 features, 15 cross-tree constraints, and 26,256 valid configurations requires 182 days. 
As a comparison, configurable systems from real-world applications such as \busybox{~}\footnote{\busyboxLink} typically consist of more than 631 features and 1,312 cross-tree constraints, allowing more than 13,402 valid configuration options. 
Therefore, the trade-off between testing time and system coverage must also consider the number of configurations for testing. 

Sample-based testing counteracts the challenge posed by the combinatorial explosion problem by generating a small but representative subset (i.e., a sample) of all possible configurations for testing~\cite{LKL:SPLC12, MKR+:ECOOP15, VAT+:SPLC18}.
A promising criterion to find a representative subset of configurations is to cover all possible combinations of feature tuples for size \twiseText (i.e., achieving t-wise feature interaction coverage) uniformly for all features~\cite{CDS:TSE08, MGSH:SPLC13, VAT+:SPLC18}. 
%For instance, a feature interaction coverage of size $t = 2$ (pair-wise coverage) is achieved when all combinations of all 2-tuples (t-tuples) of features are included in at least one configuration of the generated sample. 
Modern sampling algorithms generate samples that achieve t-wise feature interaction coverage in a short time~\cite{VAT+:SPLC18}. 
However, many algorithms only scale to small values of t ($t \leq 3$) or generate samples that are still too large for testing configurable systems in the available time~\cite{KTS+:VaMoS20}.
For example, the sampling algorithm YASA~\cite{KTS+:VaMoS20} calculates a sample that achieves three-wise feature interaction coverage of size 196 for BusyBox in about 61 minutes. %\todoTP{start 11:06 end 12:07}. 
The YASA sampling algorithm dramatically reduces the number of configurations for testing.
However, the resulting sample is still not small enough to make sample-based testing for frequently evolving configurable systems feasible.

Recent sampling approaches attempt to adapt t-wise feature interaction sampling to meet the demands of frequently evolving systems~\cite{OBM+:FSE17, MOP+:SPLC19, LSZ+:SPLC23, BBG+:SPLC23, PHK+:SPLC23, ATM+:SPLC14subsumedbyATL+:SoSyM19, PES+:MASE20, LB+:VaMoS17}.
Many approaches soften the requirement to achieve full t-wise feature interaction coverage by applying random sampling~\cite{OBM+:FSE17, MOP+:SPLC19}.
Other approaches try to utilize the evolution of configurable systems to achieve t-wise feature interaction coverage incrementally over time and, therefore, lessen the test effort for each system version~\cite{BBG+:SPLC23, PHK+:SPLC23}. 
Again, other approaches utilize the criticality of features to prioritize configurations for testing but do not guarantee t-wise feature interaction coverage of the tested set of configurations~\cite{PES+:MASE20, LB+:VaMoS17, ATM+:SPLC14subsumedbyATL+:SoSyM19}.
All of these approaches provide benefits but also have weaknesses.
For instance, random and incremental sampling approaches do not guarantee that critical features are covered with enough t-wise feature interaction coverage for every system version.
Prioritization approaches often do not ensure a certain degree of t-wise feature interaction coverage.

In this paper, we contribute to the ongoing research of adapting t-wise interaction sampling to the requirements of frequently evolving systems by introducing \mulTiWise{}. 
Our approach categorizes features into various subsets based on their criticality and covers each subset with individual strengths of t-wise feature interaction coverage.
The categorization of features into subsets is independent of measuring the criticality of a feature, which means that our approach supports various metrics to determine the criticality of features (e.g., risk assessments~\cite{LB+:VaMoS17, PES+:MASE20, Aml+:JSS00}, and change impact analysis~\cite{PES+:MASE20, MiW+:2011SAC11}). 
We combine the concepts of feature prioritization and systematical t-wise feature interaction sampling into an algorithm to mitigate the weaknesses of both approaches and strengthen their benefits.
Compared to existing approaches, \mulTiWise{} reduces the number of system configurations for testing while still covering critical feature interactions. %with high enough coverage. 

We evaluate our approach by applying \mulTiWise{} to four subject systems (i.e., \busybox{~}\footnote{\busyboxLink}, \fiasco{~}\footnote{\fiascoLink}, \soletta{~}\footnote{\solettaLink}, \uclibc{~}\footnote{\uclibcLink}) from real-world applications and compare the resulting sample sizes and the time to generate a sample against a state-of-the-art t-wise sampling algorithm. 
%We also analyze the achieved t-wise feature interaction coverage for critical feature interaction in the sample generated by adaptive sampling and state-of-the-art sampling algorithms.
Our results indicate that we can reduce the number of configurations depending on the number of critical features and the degree of t-wise feature interaction coverage with which they are covered. 
If many features are critical and need to be covered with high t-wise feature interaction coverage, we do not see much reduction in the sample sizes.

In summary, we make the following contributions to improving sample-based testing: 
	\begin{itemize}
		\item We propose \mulTiWise{} a novel approach to systematically cover subsets of features with different strengths of t-wise feature interaction coverage.
		\item We provide an open-source implementation of a sampling algorithm that utilizes our concept \footnote{\footReplication}
		\item We evaluate our concept on four real-world configurable systems \footnote{\footResults}
	\end{itemize}
%
%Foundations
\section{Foundations}
\label{section:background}
In this section, we describe the foundations to understand the context of this paper and the concepts presented later. 
%Using a simplified automotive system with 11 configuration options as a running example, we illustrate existing approaches and necessary background knowledge.
%\autoref{fig:runningexample} visualizes the hierarchical structure between these 11 configuration options as feature diagram~\cite{ABKS13, KCH+:90}. 

%% Configurable System
\begin{figure}
	\centering
	\includegraphics[width=0.8\linewidth]{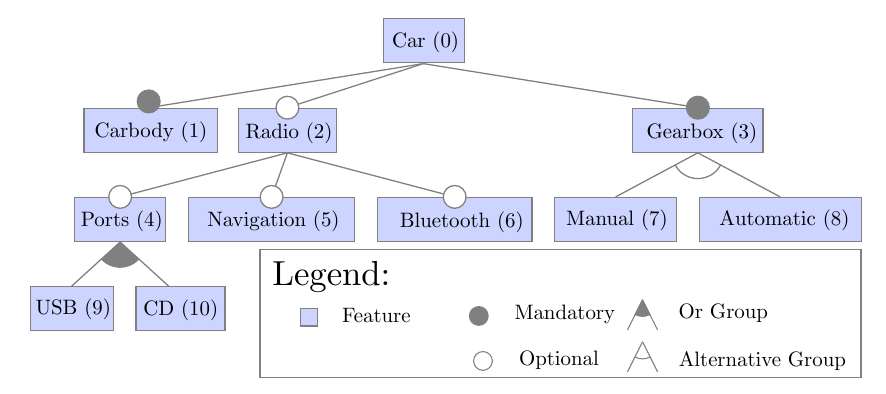}
	\caption[Car Feature Diagram]{Feature diagram of a simplified car system consisting of 11 features. The feature boxes in the diagram show the feature's name and its literal representation in parentheses.}
	\label{fig:runningexample}
\end{figure}

%________________________________________________________________________________________
\subsection{Feature Modelling}
\label{subsec:manageVariability}
\paragraph*{Basic of Feature Modelling}
A configurable system consists of products that share common core properties but differ in variable configuration options~\cite{ABKS13}. 
Customers select and deselect configuration options to customize their final product from the configurable system. 
Variability Models are typically used to express the variability of a configurable system by capturing the configuration options and their dependencies. 
According to~\cite{PHK+:SPLC23}, we define a variability model $\model = (\featureSpaceLHS, \depSpaceLHS)$ as a tuple consisting of the set of all configuration options $\featureSpaceLHS = \featureSpaceRHS$ (i.e., features) and the set of all dependencies between them $\depSpaceLHS = \depSpaceRHS$. 
Feature diagrams visualize the features and dependencies of variability models as a hierarchical tree structure. 

\autoref{fig:runningexample}, shows the feature diagram of a simplified automotive system, which consists of the root feature \texttt{Car} and its ten child features.
The features \texttt{Carbody} and \texttt{Gearbox} are mandatory child features of \texttt{Car}, which means that they must be selected if their parent is selected. 
The feature \texttt{Radio} is an optional child feature of \texttt{Car}, meaning it can either be selected or deselected if its parent is selected.
The features \texttt{Ports}, \texttt{Navigation}, \texttt{Bluetooth} are optional child features of Radio. 
The \texttt{Ports} feature has two child features \texttt{USB} and \texttt{CD} in an OR-Group, meaning that if \texttt{Ports} is selected, at least one of the child features must be selected. 
The features \texttt{Manual} and \texttt{Automatic} are child features of \texttt{Gearbox} contained in an Alternative-Group, which means that exactly one of the features must be selected if their parent is selected.
Typically, the feature diagram contains cross-tree constraints to model non-hierarchical dependencies between features of the variability model. 
Cross-tree constraints are visualized as a logical formula below the feature diagram. 
In the case of our running example, no cross-tree constraints exist. 

Another way of representing a variability model is to use a logical formula in CNF format. 
Using this representation, each clause in the CNF formula represents a dependency between features. 
For instance, the clause $(\text{\texttt{Car}} \implies \text{\texttt{Carbody}})$ represents the mandatory parent-child dependency between the feature \texttt{Car} and \texttt{Carbody}. 
Visualizing the variability model in CNF notation requires a more space-efficient representation of features than referring to their feature name. 
Typically, a shorthand notation using literals (i.e., integer values) is used for this purpose~\cite{KTS+:VaMoS20, Kri+:SPLC20}.
We define the set of all literals for a feature model $\model$ as $\litSpaceLHS = \litSpaceRHS$, where the number of existing literals equals the total number of features in the feature model. 
In our running example, we assign feature names to literals by counting the features from the top left of the feature model to the bottom right so that literal $\literal_0 = 0$ represents the \texttt{Car} feature and $\literal_{10} = 10$ represents feature \texttt{CD}. 
\autoref{fig:runningexample} indicates the literal assignment for our running example by showing the literal representing a feature in brackets beside the feature name. %(e.g., \texttt{Car (0)}).

\paragraph*{Configurations of a Configurable System}
The variability model of a configurable system represents all product configurations $\confSpace = \{\conf_1, \conf_2, \dots, \conf_n\}$ that can be derived from the system by selecting and deselecting features~\cite{ABKS13, CN01, PBL05}.
We define a complete configuration $\conf = (\fsel, \fdes)$ from the configuration space $\confSpace$ as a pair containing the set of selected features ($\fsel \subseteq \featureSpaceLHS$) and the set of deselected features ($\fdes \subseteq \featureSpaceLHS$).
We require that the sets of selected and deselected features are disjunct $\fsel \cap \fdes = \emptyset$, which means that the intersection between both sets results in the empty set.
We also require that the union of selected and deselected features results in the set of all features in the feature model ($\featureSpaceLHS = \fsel \cup \fdes$).
We use a shorthand notation using literals instead of feature names to express configurations and express deselected features by the negation operator ($\neg$). 
For instance, a minimal configuration for our running example where features \texttt{Car}, \texttt{Carbody}, \texttt{Gearbox}, and \texttt{Manual} are selected is expressed in shorthand notation as $\conf_{example} = \setOpen 0,1,3,7, \neg 2, \neg 4, \neg 5, \neg 6, \neg 8, \neg 9, \neg 10$.
A configuration is \emph{valid} as long as all dependencies defined by the set of dependencies $\depSpaceLHS$ from the feature model can be fulfilled by the feature selection of the configuration.
For instance, our example configuration is valid because the requirement of the alternative group under \texttt{Gearbox} is fulfilled by $\conf_{example}$.
Selecting the feature \texttt{Automatic} (7) in addition to the feature \texttt{Manual} leads to an invalid configuration because the feature selection contradicts the alternative group below feature \texttt{Gearbox}.
%------------------------------
\subsection{Configuration Testing}
\label{subsec:confTesting}
Product-based testing uses valid configurations to assure the functional safety of configurable systems. 
However, in practice, testing all valid configurations of the configurable system is often not feasible because of the combinatorial explosion problem~\cite{HNA+:EMSE19}. 
Various approaches exist to select a representative subset (i.e., a Sample) of configurations for testing~\cite{MK+:ICSE16, VAT+:SPLC18}. 
%We define a sample $\samp = \setOpen \conf_1, \conf_2, \dots \conf_n \setClose | \samp \subseteq \confSpace_{valid}$ as subset of all valid configurations for a feature model $\model$.
%Uniform random sampling~\cite{OGB+:TR20, HS+:VaMoS24, BaL+:SPLC22} is a commonly used strategy to select configurations for testing because it generates an unbiased representation of valid system configurations.
%However, uniform random sampling does not provide system coverage criteria to guarantee the quality of samples. 
T-wise feature interaction sampling is one of those approaches that consider the coverage of all valid combinations of t-wise feature tuples ($\tTuple$) as a quality criterion for samples~\cite{CDS+:ISSTA07, JHF+:MODELS12, KTS+:VaMoS20}.
A t-wise feature tuple, is a tuple ($\tTuple = (\literal_1, \literal_2, \dots, \literal_t)$) of size \twiseText that contains features from a feature model.
For instance, $(7, 8)$ is a pair-wise feature tuple for the features \texttt{Manual} and \texttt{Automatic} from our running example.

Building all permutations of selected and deselected features in the t-wise feature tuple generates all possible t-wise feature interactions for the t-wise feature tuple. 
We define a t-wise feature interaction $\interactionLHS = \interactionRHS$ as a tuple of size \twiseText that contains selected and deselected features from the feature model $\model$. 
For instance, $\setOpen (7,8), (\neg 7, 8), (7, \neg 8), (\neg 7, \neg 8) \setClose$ are all feature interactions for the pair-wise ($\twise = 2$) feature tuple $\tupleOpen 7, 8 \tupleClose$
%for the features \texttt{Manual} and \texttt{Automatic} from our running example.
A feature interaction of size \twiseText is valid when it can appear in at least one configuration of the configuration space $\confSpace$ for a feature model.
The pair-wise feature combinations $(\neg 7, 8), (7, \neg 8)$ are valid for our running example, while the feature combinations $(7, 8), (\neg 7, \neg 8)$ are not. 
%The combinations $(7, 8), (\neg 7, \neg 8)$ contradict the alternative group below the feature \texttt{Gearbox}, which states that exactly one of the features \texttt{Manual} or \texttt{Automatic} must be selected but not both, therefore, those feature interactions are invalid.
We define the set of all valid \twiseText feature interactions $\setOpen I_{1}, \dots, I_{n} \setClose$ for a set of features %$\featureSpaceLHS$ in a feature model $\model$ 
as $\intersetLHS{t}{\model}{\featureSpaceLHS} = \intersetRHS$.

A t-wise feature interaction is covered by a configuration in a sample when the interaction tuple is contained in at least one configuration of the sample. 
We define the set of t-wise feature interactions in a sample by $\interInSamLHS = \interInSamRHS$.
A sample achieves full (100\%) t-wise feature interaction coverage when all t-wise feature interactions are covered by at least one configuration of the sample.
Accordingly, we define the ratio of t-wise feature interaction coverage that a sample achieves by dividing the number of all valid feature interactions in the sample by the number of all valid feature interactions of the feature model. 
%\autoref*{eq:coverage} shows the resulting formula.

%\begin{equation}
%	\label{eq:coverage}
%	\osCoverLHS = \osCoverRHS
%\end{equation}

%
% MulTiWise Sampling
\section{\mulTiWise{}}
\subsection{Problemstatement}
T-wise feature interaction coverage is a prominent metric to rate the effectiveness of samples for testing product lines\cite{VAT+:SPLC18}. 
%Many software failures are caused by the interaction of selected and deselected features rather than the presence or absence of a single feature. 
Testing a set of configurations that achieve full (100\%) t-wise feature interaction coverage for high values of \twiseText (i.e., $\twise = 3$, $\twise = 4$, $\twise = 5$, etc.) promises a high chance of discovering a fault in a product line. 
Modern sampling algorithms such as \yasa{} can calculate samples that achieve t-wise feature interaction coverage for various values of \twiseText (e.g., $\twise = 1$, $\twise = 2$, $\twise = 3$, $\twise = 4$, etc.)~\cite{KTS+:VaMoS20}. 
However, generating samples that achieve full t-wise feature interaction coverage for higher values of \twiseText (i.e., $\twise > 2$) becomes more time-consuming and results in more configurations with an increasing t-value because exponentially more feature interactions must be considered when generating a sample.
For instance, using the \yasa{} sampling algorithm to generate a pair-wise ($\twise = 2$) sample for our running example takes only a few milliseconds, and the resulting sample contains seven configurations.
In contrast, calculating a sample that achieves three-wise feature interaction coverage ($\twise = 3$) takes three seconds, and the resulting sample contains 18 configurations. 
While calculating a sample in three seconds and testing 18 configurations seems manageable, for larger systems, the number of configurations increases exponentially with the number of optional features in the feature model~\cite{HNA+:EMSE19}. 
In industry branches where testing each configuration requires lots of monetary resources (i.e., safety-critical cyber-physical systems), testing an enormous number of configurations is not feasible. 
Therefore, the challenges of increasing the time to generate samples and the increase in the number of configurations for testing limit the application of samples, achieving full (100\%) t-wise feature interaction coverage for higher values of t (i.e., $t > 2$).
%Instead, selecting configurations for testing often relies on expert knowledge, which leads to an unstructured testing process.

\subsection{Solution Idea}
In practice, certain groups of features exist, for which achieving higher t-wise feature interaction coverage is potentially more valuable than for other groups of features. 
For instance, covering the feature interactions between a group of features that share highly interconnected implementation artefacts with higher values of \twiseText is probably more valuable than doing so for features that do not share any implementation artefacts.
Other examples of more valuable feature groups include safety-critical features of a system and recently changed features.
In our running example, the features \texttt{Carbody}, \texttt{Manual}, and \texttt{Automatic} strongly interact with each other. 
Therefore, they belong to a critical feature group for which the testing requirement defines pair-wise coverage.
The features \texttt{Car}, \texttt{Radio}, and \texttt{Gearbox} are important to premium customers and therefore belong to a feature group requiring one-wise feature interaction coverage. 
For the remaining features, no special t-wise feature interaction coverage is required.

We aim to use the differences between testing requirements for feature groups to enable a dynamic tradeoff between large samples that achieve full (i.e., 100\%) t-wise feature interaction coverage for all features equally for the same \twiseText value and smaller samples that achieve full t-wise feature interaction coverage for feature groups with specially assigned \twiseText values.
%The concept behind \mulTiWise{} is to generate samples that cover the features of relevant t-wise feature interaction groups with high t-wise feature interaction coverage. 
%In contrast, features in non-relevant groups are covered with lower t-wise feature interaction coverage.
%We aim to enable a dynamic tradoff between large samples that achieve full (i.e., 100\%) t-wise feature interaction coverage for all features equally for the same \twiseText value and smaller samples that achieve full t-wise feature interaction coverage for feature groups with specially assigned \twiseText values. 
To do so, we leverage the correlation that covering fewer t-wise feature interactions results in fewer configurations for testing. 
Compared to sampling algorithms that try to achieve full t-wise feature interaction coverage for all features equally, we reduce the number of t-wise feature interactions using two approaches. 
Firstly, we split the whole feature set into multiple distinct feature groups and consider only feature interactions between features in the groups when calculating a sample. 
For instance, all valid pair-wise feature interaction between the features \texttt{Carbody}, \texttt{Manual}, and \texttt{Automatic} and all valid one-wise feature interactions between the features \texttt{Car}, \texttt{Radio}, and \texttt{Gearbox} will be actively considered when calculating a sample with \mulTiWise{}. 
However, t-wise feature interactions between the feature groups such as interactions between the features \texttt{Car}, \texttt{Manual}, \texttt{Bluetooth}, will not be considered during the sample calculation. 

The second approach to reduce the t-wise feature interactions to be covered is that each feature group gets its own \twiseText value assigned. 
Therefore, we can specify small groups of features that will be covered with high t-wise feature interaction coverage and large groups of features that will be covered with low t-wise feature interaction coverage. 
For instance, in our running example, we specify pair-wise coverage for the small group of features \texttt{Carbody}, \texttt{Manual}, and \texttt{Automatic}, one-wise coverage for another small group of features \texttt{Car}, \texttt{Radio}, and \texttt{Gearbox}, and no coverage criteria for the remaining features. 
Doing so largely reduces the feature interactions to be covered, compared to specifying a pair-wise feature interaction coverage criterion for all features equally. 
\subsection{Feature Grouping}
\mulTiWise{} uses groups of features that get a certain \twiseText value assigned as a basic concept to generate samples.
We formally define such a group of features $\tGroupLHS = \tGroupRHS | \tGroupLHS \subseteq \featureSpaceLHS$ as a set of features from the feature model $\model$. 
We define the set of all t-wise feature interaction groups as $\tGroupSetLHS = \tGroupSetRHS$. 
$\tGroupDefault$ represents a default t-wise feature interaction group that contains all features from the feature set $\featureSpaceLHS$ that are not contained in any other t-wise feature interaction group.
A feature $\feature \in \featureSpaceLHS$ can simultaneously be part of multiple t-wise feature interaction groups.
The union, $\bigcup \tGroupSetLHS = \featureSpaceLHS$ of all t-wise feature interaction groups, is equal to the set of all features defined by the feature model.
We assign a t-wise feature interaction coverage value $\twiseText$ to each $\tGroupLHS$.
Multiple t-wise feature interaction groups may have the same t-wise feature interaction value. 
%%%
%% table that shows t-wise feature interaction groups for the running example
\begin{table}
	\centering
	\caption{Assignment of features from the running example to t-wise feature interaction groups.}
	\begin{tabular}[t]{lll} 
		\toprule
		$\tGroupLHS$ 		& t-value		& Features
		\\ \toprule
		$\tGroupLHS_1$ 		& 1 			& $\setOpen \text{\texttt{Car},\texttt{Radio}, \texttt{Gearbox}} \setClose$ \\
		$\tGroupLHS_2$ 		& 2 			& $\setOpen \text{\texttt{Carbody},\texttt{Manual},\texttt{Automatic}} \setClose$ \\
		$\tGroupDefault$	& 0 			& $\setOpen \text{\texttt{USB},\texttt{CD}}, \texttt{Ports}, \texttt{Navigation}, \texttt{Bluetooth} \setClose$ \\
		\bottomrule	
	\end{tabular}
	\label{tab:groupingTable}
\end{table}

% All Features: Car, Carbody, Radio, Gearbox, Ports, Navigation, Bluetooth, Manual, Automatic, USB, CD
% TG1: Car, Radio, Gearbox
% TG2: Carbody, Manual, Automatic
% TGD: USB, CD, Ports, Navigation, Bluetooth 
%%
%%%%
\autoref{tab:groupingTable} shows an example of t-wise feature interaction groups for our running example. 
We define three t-wise feature interaction groups, of which one is the default group.
Each group gets assigned a t-wise feature interaction value ($\twise$), shown in column two of \autoref{tab:groupingTable}. 
Column three shows which features from the feature model are assigned to each group.  
For instance, in \autoref{tab:groupingTable}, we see that $\tGroupLHS_1$ contains the features \texttt{Car}, \texttt{Radio}, \texttt{Gearbox} and that this group has a t-value of $\twise = 1$ assigned. 
The features \texttt{Carbody}, \texttt{Manual}, and \texttt{Automatic} are in $\tGroupLHS_2$ with a t-value of $\twise = 2$. 
The default t-wise interaction group $\tGroupDefault$ contains the features \texttt{USB}, \texttt{CD}, \texttt{Ports}, \texttt{Navigation}, \texttt{Bluetooth}, which are not assigned to any other t-wise feature interaction group in our example. 
We assign a t-value of zero ($\twise = 0$) to the default t-wise feature interaction group, meaning that considering interactions between those features for the resulting sample is optional. 
\subsection{Generating MulTi-Wise Samples}
\label{subsec:sampling}
In \autoref{alg:adaptiveSampling} we show a pseudocode algorithm to visualize the generation process for a sample using \mulTiWise{}. 
This algorithm uses a feature model ($\model$), as well as a set of t-wise feature interaction groups ($\tGroupSetLHS$) as input to generate a sample ($\samp$). 
The resulting sample achieves t-wise feature interaction coverage for all given t-wise feature interaction groups for the respective t-value of each group.
%In the case of our running example, the resulting sample will achieve one-wise feature interaction coverage for the features in $\tGroupLHS_1$ and pair-wise feature interaction coverage for the features of $\tGroupLHS_2$. 
%Since we defined no specific t-wise coverage (i.e., a t-value of zero) for $\tGroupDefault$, our algorithm does not guarantee a certain t-wise coverage for the included features.
%%%
%% pseudocode algorithm for \mulTiWise{}
%pseudocode algorithm for adaptive sampling strategy

\SetKwComment{Comment}{/* }{ */}
\begin{algorithm}
    \caption{MulTiWise Sampling Algorithm}\label{alg:adaptiveSampling}
    \KwIn{$\model = (\featureSpaceLHS, \depSpaceLHS), \tGroupSetLHS = \tGroupSetRHS$}
    \KwData{$\samp = \emptyset, \twise = 0,$}
    \KwResult{$\samp$}

    \ForEach{$\tGroupLHS \in \tGroupSetLHS$}{
        $\twise \leftarrow \twise(\tGroupLHS)$\;
        $\samp' \leftarrow \text{\textcolor{kitGreen1}{coveringStrategy}}(\model, \tGroupLHS, \samp)$\;
        $\samp \leftarrow \samp \cup \samp'$\;
    }

\end{algorithm}
%%
%%%%
Our algorithm iterates over the set of t-wise feature interaction groups provided as input (see line 1).
As shown in line 2, the algorithm extracts the t-value of the respective t-wise interaction group. 
After that, the algorithm generates an intermediate sample $\samp'$ that achieves the respective t-wise feature interaction coverage for the features in the current feature interaction group (see line 3).
The final step of our algorithm (see line 4) merges the intermediate sample $\samp'$ with the global result sample $\samp$ by adding all configurations from $\samp'$ to $\samp$ that do not already exist in $\samp$.

We use the covering strategy of the existing \yasa{} sampling algorithm~\cite{KTS+:VaMoS20} to generate the intermediate sample $\samp'$ of \autoref{alg:adaptiveSampling}. 
\yasa{} is an efficient t-wise sampling technique that provides various options when calculating samples that achieve t-wise feature interaction coverage~\cite{KTS+:VaMoS20}
For instance, the \yasa{} algorithm can generate a sample that achieves t-wise feature interaction coverage for only a subset of features from the feature model. 
We use exactly this functionality of \yasa{} to calculate an intermediate sample for each t-wise feature interaction group. 
%
%% pseudocode algorithm for a basic sampling algorithm
%%
% pseudocode for a basic sampling algorithm
\SetKwComment{Comment}{/* }{ */}

\begin{algorithm}
\caption{Basic Sampling Algorithm~\cite{KTS+:VaMoS20} } \label{alg:basicSampling}
\KwIn{ $\model = \tupleOpen \featureSpaceLHS, \depSpaceLHS \tupleClose$, $t \geq 0$, $\tGroupLHS = \setOpen \literal_{1}, \dots, \literal_{n} \setClose$,  $\samp = \setOpen \conf_{1}, \dots, \conf_{n} \setClose$
}

\KwData{$\intersetLHS{t}{\model}{\tGroupLHS} = \emptyset$
}

\KwResult{$\samp = \emptyset$
}

$\intersetLHSNoParam \leftarrow \intersetLHS{t}{\model}{\tGroupLHS}$ \;
\ForEach{$\interactionLHS \in \intersetLHS{t}{\model}{\tGroupLHS}$}{
      \If{$\not \exists \conf \in \samp : \interactionLHS \in \conf$}{
        \ForEach{$\conf \in \samp$}{
          $\conf' \leftarrow \conf \cup \interactionLHS$\; 
          \If{$\text{valid}(\conf', \model)$}{
            $\samp \leftarrow (\samp \setminus \setOpen \conf \setClose) \cup \setOpen \conf' \setClose$ \;
            \Return{$\samp$}\;
            }
          }
       } 
    $\samp \leftarrow \samp \cup \interactionLHS$\;
}
$\samp \leftarrow \text{completeConfigurations}(\samp, \model)$ \;
\Return{$\samp$}\;
%\ForEach{$I \in \intersetLHS{t}{\model}{TG}$}{do something}

%$y \gets 1$\;
%$X \gets x$\;
%$N \gets n$\;
%\While{$N \neq 0$}{
%  \eIf{$N$ is even}{
%    $X \gets X \times X$\;
%    $N \gets \frac{N}{2}$ \Comment*[r]{This is a comment}
%  }{\If{$N$ is odd}{
%      $y \gets y \times X$\;
%      $N \gets N - 1$\;
%    }
%  }
%}
\end{algorithm}
We present the basic covering strategy of the \yasa{} sampling algorithm in \autoref{alg:basicSampling}. 
The algorithm takes the feature model $\model$, a t-wise coverage value $\twise$, and a set of features $\tGroupLHS$ as input and generates a sample $\samp$ as output. 
The algorithm starts by generating a set of all valid feature interaction tuples $\intersetLHS{\twise}{\model}{\tGroupLHS}$ of size $\twise$ for the feature set $\tGroupLHS$ (see line 1). 
%% table showing the interaction sets for the TGroups from the TGroup example table
%%
%% example for the interaction sets for the t-groups defined in table tgroupExample.tex
%\begin{table}
%	\centering
%	\caption{Valid feature t-wise interactions for the t-wise feature interaction groups from \autoref{tab:groupingTable}.}
%	\begin{tabular}[t]{ll} 
%		\toprule
%		$\tGroupLHS$ 		& $\intersetLHS{\twise}{\model}{\tGroupLHS}$
%		\\ \toprule
%		$\tGroupLHS_1$ 		& $\setOpen \tupleOpen \texttt{Car} \tupleClose, \tupleOpen \texttt{Gearbox} \tupleClose, \tupleOpen \texttt{Radio} \tupleClose, \tupleOpen \neg\texttt{Radio} \tupleClose\setClose$ \\
%		$\tGroupLHS_2$ 		&  $\setOpen \tupleOpen \texttt{Carbody},\texttt{Manual}  \tupleClose, \tupleOpen \texttt{Carbody}, \neg\texttt{Manual} \tupleClose,$ \\
%                            & $\tupleOpen \texttt{Carbody},\texttt{Automatic} \tupleClose, \tupleOpen \texttt{Carbody},\neg\texttt{Automatic} \tupleClose,$ \\
%                            & $\tupleOpen \texttt{Manual},\neg\texttt{Automatic} \tupleClose, \tupleOpen \neg\texttt{Manual},\texttt{Automatic} \tupleClose\setClose$ \\
%		$\tGroupDefault$	&  $\emptyset$ \\
%		\bottomrule	
%	\end{tabular}
%	\label{tab:interactionTable}
%\end{table}

\begin{table}
	\centering
	\caption{Valid feature t-wise interactions for the t-wise feature interaction groups from \autoref{tab:groupingTable}.}
	\begin{tabular}[t]{ll} 
		\toprule
		$\tGroupLHS$ 		& $\intersetLHS{\twise}{\model}{\tGroupLHS}$
		\\ \toprule
		$\tGroupLHS_1$ 		& $\setOpen \tupleOpen 0 \tupleClose, \tupleOpen 3 \tupleClose, \tupleOpen 2 \tupleClose, \tupleOpen \neg 2 \tupleClose\setClose$ \\
		$\tGroupLHS_2$ 		&  $\setOpen \tupleOpen 1, 7  \tupleClose, \tupleOpen 1, \neg 7 \tupleClose, \tupleOpen 1, 8 \tupleClose, \tupleOpen 1,\neg 8 \tupleClose, \tupleOpen 7,\neg 8 \tupleClose, \tupleOpen \neg 7, 8 \tupleClose\setClose$ \\
		$\tGroupDefault$	&  $\emptyset$ \\
		\bottomrule	
	\end{tabular}
	\label{tab:interactionTable}
\end{table}
\autoref{tab:interactionTable} shows the valid interaction tuples for the t-wise feature interaction groups presented in \autoref{tab:groupingTable}. 
We present the features of our running example in their literal notation to keep a concise representation. 
Generating feature interaction tuples for $\tGroupLHS_1$ results in four feature tuples of size one. 
The features \texttt{Car} (i.e., 1) and \texttt{Gearbox} (i.e., 3) are mandatory in our running example. 
Therefore, the feature interaction tuples $\neg 1$ and $\neg 3$ are excluded from \autoref{tab:interactionTable}. 
Two one-wise feature interaction tuples (e.g., $2$ and $\neg 2$) are valid For feature \texttt{Radio} because it is an optional feature in our running example.
\autoref{alg:basicSampling} generates six valid pair-wise feature interaction tuples for $\tGroupLHS_2$, resulting from the specified t-value $\twise = 2$ and the constraints of the feature model. 
The set of generated feature interaction tuples for $\tGroupDefault$ is empty because our example specifies a t-value of $\twise = 0$ for this group, which means that no specific t-wise feature interactions between the features in this group must be considered when generating a sample.

After generating the t-wise feature interaction tuples, the algorithm iterates over all tuples (see line 2) and checks for each interaction tuple $\interactionLHS_{n}$ whether this tuple is already covered in any configuration of the sample $\samp$. 
The algorithm proceeds with the next interaction tuple $\interactionLHS_{n+1}$ if a configuration contains the interaction tuple $\interactionLHS_{n}$.
Otherwise, it iterates over the set of existing configurations in $\samp$ and tries adding $\interactionLHS_{n}$ to the currently selected configuration $\conf'$ so that the result is still valid with regard to the feature model (see lines 3 to 5). 
If the resulting configuration $\conf'$ is still valid, $\conf'$ replaces the original configuration $\conf$ in the sample, and the algorithm continues with the next interaction tuple (see lines 6 to 9). 
Otherwise, the interaction tuple $\interactionLHS$ is added to the sample $\samp$ as a new configuration (see line 13). 
Finally, the algorithm completes each configuration in $\samp$ by selecting or deselecting all undecided feature options. 
%so that the partial configurations in $\samp$ become complete and valid configurations with regard to the feature model $\model$. 
%% table showing the results from applying \mulTiWise{} to the example feature interaction groups
%%
%% Example result for creating a sample with adaptive sampling for the running example
\begin{table}
	\centering
	\caption{Resulting sample from using \mulTiWise{}.}
	\begin{tabular}[t]{lll} 
		\toprule
				& $\conf$		& Feature Literals
		\\ \toprule
		$\text{After} \tGroupLHS_1$ 		& $\conf_{1}$ &  $\setOpen 0, 1, \neg 2, 3, \neg 4, \neg 5, \neg 6, 7, \neg 8, \neg 9,\neg 10 \setClose$ \\
											& $\conf_{2}$ & $\setOpen 0, 1, 2, 3, \neg 4, \neg 5, \neg 6, \neg 7, 8, \neg 9,\neg 10 \setClose$
        \\ \hline 
		$\text{After} \tGroupLHS_2$ 		& $\conf_{1}$ &  $\setOpen 0, 1, \neg 2, 3,\neg 4, \neg 5, \neg 6, 7, \neg 8, \neg 9,\neg 10 \setClose$ \\
											& $\conf_{2}$ & $\setOpen 0, 1, 2, 3, \neg 4, \neg 5, \neg 6, \neg 7, 8, \neg 9,\neg 10 \setClose$
                                    
        \\ \hline
        $\text{After} \tGroupDefault$	    & $\conf_{1}$ &  $\setOpen 0, 1, \neg 2, 3,\neg 4, \neg 5, \neg 6, 7, \neg 8, \neg 9,\neg 10 \setClose$ \\
        & $\conf_{2}$ & $\setOpen 0, 1, 2, 3, \neg 4, \neg 5, \neg 6, \neg 7, 8, \neg 9,\neg 10 \setClose$ \\
		\bottomrule	
	\end{tabular}
	\label{tab:exampleSample}
\end{table}

In \autoref{tab:exampleSample}, we show the stepwise process of generating a sample with \mulTiWise{}. 
As we show in \autoref{alg:adaptiveSampling}, \mulTiWise{} is an iterative process that generates an intermediate sample for t-wise feature interaction groups and uses them further in the sampling process.
In the case of our running example, \mulTiWise{} starts with calculating an intermediate sample for $\tGroupLHS_1$. 
Covering all feature tuples for $\tGroupLHS_1$ (see the first row of \autoref{tab:interactionTable}) requires two configurations (visualized with green colour in \autoref{tab:exampleSample}). 
%For our running example, we complete partial configurations by deselecting as many features as possible to have a deterministic and comprehensible approach to completing partial configurations.
%In practice, more sophisticated completion strategies can reduce the total number of configurations for large systems.
After the sample for $\tGroupLHS_1$ is generated, we generate an intermediate sample that covers the feature interactions identified for $\tGroupLHS_2$ (see the second row of \autoref{tab:exampleSample}).
We use the intermediate sample generated for $\tGroupLHS_1$ as input. 
The previously generated configurations $\conf_1$ and $\conf_2$ already cover three of the six feature interaction tuples for $\tGroupLHS_2$ (visualized with green colour). 
Our procedure generates a third configuration to cover the remaining feature interaction tuples of $\tGroupLHS_2$ to finalize the intermediate sample for this iteration. 
The default t-wise feature interaction group $\tGroupDefault$ of our running example does not introduce any t-wise feature interaction tuple that needs to be covered by \mulTiWise{}. 
Therefore, our procedure generates no more configurations and returns a final sample containing the configuration $\conf_1$, $\conf_2$, and $\conf_3$ visualized in the last row of \autoref{tab:exampleSample}.
\label{sec:concept}
%
% Evaluation
\section{Evaluation}
%In our evaluation, we investigate whether \mulTiWise{} can reduce the sample size and the time needed to generate a sample (sampling time) while still achieving feature interaction coverage. 
In our evaluation, we investigate whether \mulTiWise{} enables a tradeoff between large samples that achieve full t-wise feature interaction coverage for all features equally and small samples that achieve t-wise feature interaction coverage for specified groups of features.
Our investigation focuses on the \emph{robustness} and \emph{performance} of our approach to show that \mulTiWise{} is feasible in practice.
We analyze these aspects of our approach by answering the following research questions. 

%\paragraph{\textbf{(Feasibility and Correctness)} RQ1: Does \mulTiWise{} achieve the required t-wise feature interaction coverage for all features marked with the associated t-value?}
%\label{rq1}
%In RQ1, we investigate the influence of explicitly covering groups of features with different t-wise interaction coverage values (t-values). 
%As part of the experiment, we assign each feature of a feature model to a feature-interaction group with a certain t-value and perform \mulTiWise{} as described in \autoref{subsec:sampling}. 
%We measure the achieved t-wise feature interaction coverage on the resulting sample for different t-values in percent, considering all features of the feature model. 
%In addition, we analyze the percentage of covered feature interactions defined by the different feature groups.
%We expect that, for each feature group, the sample covers all t-wise feature interactions defined by the respective feature group. 
%We also expect that the resulting sample will cover a higher percentage of t-wise feature interactions for t-values lower than the highest t-value defined by a feature group. 
%The closer the t-value gets to the highest t-value defined by a feature group, the smaller the percentage of covered t-wise feature interactions will get.

\paragraph{\textbf{(Robustness)} RQ1: How does the size of feature groups influence sampling metrics of \mulTiWise{}? }
\label{rq1}
In RQ1, we investigate how changing the size of feature groups with assigned t-values influences sampling metrics such as sample size, sampling time, and achieved percentage of t-wise feature interaction coverage.
In particular, we are interested in investigating how gradual changes in the size between two feature interaction groups with different \twiseText values influence our sampling metrics.
Therefore we specify multiple experiment setups, where we distirbute all features from a subject system between two t-wise feature interaction groups. 
In the extreme cases of those experiment setups, all features (i.e., 100\%) are assigned to only one t-wise feature interaction group. 
We define the other experiment setups by gradually distributing a percentage of features from one t-wise feature interaction group to the other.
We execute \mulTiWise{} for all experiment setups and measure sample size, sampling time, and achieved t-wise feature interaction coverage.
%the extreme cases of size deviations between feature groups. 
%We start our experiment for this research question by uniformly distributing all features of the feature model to different t-values (i.e., each t-value in the experiment gets assigned the same number of features). 
%We measure the sample size, sampling time, and the achieved t-wise feature interaction coverage for different values of \twiseText as the baseline for this experiment. 
%Next, we create permutations of our baseline by shifting the number of features between the t-value groups. 
%We measure the sampling metrics again for each permutation and compare them against the baseline. 
We assume that the sample size, the sampling time, and the achieved t-wise feature interaction coverage will increase when more features are assigned to a t-wise feature interaction group with a higher \twiseText value. 
%On the contrary, we expect all three metrics to decrease when many features are assigned to smaller t-values.

\paragraph{\textbf{(Performance)} RQ2: How does \mulTiWise{} compare to state-of-the-art t-wise sampling algorithms?}
\label{rq2}
With RQ2, we investigate the performance of \mulTiWise{} compared to state-of-the-art sampling algorithms. 
%In particular, we want to analyze whether \mulTiWise{} reduces the sampling size of generated samples while achieving t-wise feature interaction coverage values comparable to those of state-of-the-art sampling algorithms.
%As the baseline of our comparison, we generate samples using the \yasa~\cite{KTS+:VaMoS20} sampling algorithm. 
In particular, we are interested in how changing the size of t-wise feature interaction groups changes the sample size, the sampling time, and the achieved t-wise feature interaction coverage compared to a baseline algorithm.
We choose the \yasa sampling algorithm as the baseline for our comparison because in previous evaluations~\cite{KTS+:VaMoS20} it shows the best performance values compared to other sampling algorithms proposed in the literature. 
\yasa generates samples that achieve t-wise feature interaction coverage for all features equally, which is the same as assigning all features to only one feature interaction group in our \mulTiWise{} setting (i.e., the extreme cases from RQ1).
We execute \yasa for the two \twiseText values assigned to the t-wise feature interaction groups from RQ1, and measure the sample size, the sampling time and the achieved t-wise feature interaction coverage.
We then compare the measured results with those measured in the experiments for RQ1. 
We assume that \mulTiWise{} achieves comparable results to \yasa for the extreme setups of our experiments. 
Therefore we also assume, that the findings for the experiment setups in between the extreme cases will be comparable to those of RQ1. 
%Therefore, we chose \yasa as the base for our comparison. 
%Our experiment uses \yasa to generate samples that achieve full t-wise feature interaction coverage for various t-values. 
%We track the sample size, sampling time, and the percentage of achieved t-wise feature interaction coverage for each generated sample for various t-values.
%We then compare the results of \mulTiWise{} acquired in the experiment from \autoref{rq2} and the results from generating samples using \yasa with each other. 
%We then compare the results of \mulTiWise{} acquired in the experiment from \autoref{rq1} and the results from generating samples using \yasa with each other.
%We assume that the samples generated with \mulTiWise{} will not achieve complete (100\%) t-wise feature interaction coverage for all t-values in our experiment. 
%In contrast, the samples generated by \yasa will achieve full t-wise feature interaction coverage for the respective t-values. 
%However, we expect samples generated with \mulTiWise{} to be smaller than those generated with the \yasa sampling algorithm. 
%We also expect the time to generate a sample to be comparable between both algorithms. 

\subsection{Experiment Setup}
\label{subsec:ExSet}
%In this section, we define the general framework of how we execute our experiments.  
%This includes defining the subject systems we will use in the experiment. 
%We also describe how we assign feature attributes to the subject systems. 
%Finally, we describe the hardware used to perform the experiments.
%
\paragraph{Subject Systems}
\label{par:exSet:subjectsys}
We perform our evaluation on the subject systems \listSubjectsystems~\footnote{\listSubjectsystemsFoot}. 
Pett et al.~\cite{PHK+:SPLC23} already used extensive feature model histories of those subjects to evaluate their \emph{Continuous T-Wise Sampling} approach.
In our experiment, we do not need to analyze an extensive feature model history but only the most recent feature model of each subject system.
%Therefore, our experiment uses only each subject system's most recent feature model.
%
\begin{table}
	\centering
	\caption{Overview Subject Systems}
	\begin{tabular}[t]{lll} 
		\toprule
		Subject System 		& Features					& Constraints
		\\ \toprule
		\busybox{} 			& 631  	&  1312 \\
		\fiasco{} 				& 253 	&  1795 \\
		\soletta{} 			& 457 	&  2319 \\
		\uclibc{} 			& 235 	&  1905 \\
		\bottomrule	
	\end{tabular}
	\label{tab:subjectsystemTable}
\end{table}
\autoref{tab:subjectsystemTable} shows the number of features and the number of constraints for the most recent feature models in our subject systems' available feature model history.
Those values reveal that our subject systems vary from a large feature model with over 631 features and 1312 constraints to medium-sized feature models with about 253 features and 1795 constraints. 
\paragraph{Experiment Setups}
\label{par:exSet:exSet}
%Performing experiments for all possible t-wise feature interaction groups is not feasible.
%Therefore, we evaluate \mulTiWise{} with a static subset of two feature interaction groups with the t-values two and three.
We evaluate \mulTiWise{} with a static subset of two feature interaction groups with the t-values two and three.  
In doing so, our experiments include a group with a lower t-value (i.e., t=2) and one group with a higher t-value (i.e., t=3 ). 
%
%%%%
\begin{table}
    \caption{Experiment Setups.}
    \centering
    \begin{tabular}{ c c c c c }
    \toprule
    Experiments   & $\tGroupLHS$ & t-value & percentage of features \\
    \toprule
    \rowcolor{kitWhite3}
    Exp1          & \multicolumn{3}{c}{Pair-Wise YASA Sampling ($\twise = 2$)}\\ \hline
    Exp2          & $\tGroupLHS_{t2}$ & 2       & 100     \\
                  & $\tGroupLHS_{t3}$ & 3       & 0       \\ \hline
    Exp3          & $\tGroupLHS_{t2}$ & 2       & 75      \\
                  & $\tGroupLHS_{t3}$ & 3       & 25      \\ \hline
    Exp4          & $\tGroupLHS_{t2}$ & 2       & 50      \\
                  & $\tGroupLHS_{t3}$ & 3       & 50      \\ \hline
    Exp5          & $\tGroupLHS_{t2}$ & 2       & 25      \\
                  & $\tGroupLHS_{t3}$ & 3       & 75      \\ \hline
    Exp6          & $\tGroupLHS_{t2}$ & 2       & 0       \\
                  & $\tGroupLHS_{t3}$ & 3       & 100     \\ \hline 
    \rowcolor{kitWhite3}
    Exp7          & \multicolumn{3}{c}{Three-Wise YASA Sampling ($\twise = 3$)} \\
    \bottomrule
    \end{tabular}
    \label{tab:experimentSetup}
\end{table}
%%%%
%
\autoref{tab:experimentSetup} shows seven experiment setups, from which two are baseline setups (visualized by grey background), and five are setups for \mulTiWise{}.
As baseline setups, we use two setups of the \yasa sampling algorithm one to achieve full pair-wise feature interaction coverage (i.e., \setupOne{}) and one to achieve full three-wise feature interaction coverage (i.e., \setupSev{}).
The experiment setups of \mulTiWise{} define two t-wise feature interaction groups, one with a t-value of two (i.e., $\tGroupLHS_{t2}$) and one with a t-value of three (i.e., $\tGroupLHS_{t3}$). 
%Through all of the setups, the group definition does not change. 
The varying factor between the setups is the number of features assigned to each t-wise feature interaction group.
Column four of \autoref{tab:experimentSetup} shows the feature distributions per t-wise feature interaction group in per cent.
The first experiment setup of \mulTiWise{} (i.e., \setupTwo{}) assigns 100\% of the features to feature group $\tGroupLHS_{t2}$ and 0\% of the features to feature group $\tGroupLHS_{t3}$.
The following experiment setups (i.e., \setupThree{} to \setupSix{}), reduce the number of features assigned to $\tGroupLHS_{t2}$ by 25\% while increasing the number assigned to $\tGroupLHS_{t3}$ by the same percentage, until \setupSix{} assigns 0\% of features to $\tGroupLHS_{t2}$ and 100\% to $\tGroupLHS_{t3}$.
\paragraph{Evaluation Hardware}
We implement \mulTiWise{} as a Java command line tool. 
The tooling integrates functionality from the FeatureIDE library\footnote{\footFIDE}.
All of our tooling is available as part of our replication package\footnote{\footReplication}
We execute the tooling on a virtual server running Ubuntu 20.04 as an operating system and with an OpenJDK version 1.1.8.0\_292 as the running version of Java. 
The server has a processor with eight cores running at 2400MHz and is equipped with 16GB of physical memory, from which we use 12GB as virtual memory to execute our experiments.
%\begin{remark}
For the review version of this paper, we cannot provide the source code of our tooling because it may contain hints of the author's identities, which contradicts the double-blind review regulations. 
We will include the source as part of the replication package in the final version of this paper.
%\end{remark}
%
\subsection{Experiment Execution}
\label{subsec:exex}
In \autoref*{tab:experimentSetup} we specify two baseline experiment setups (i.e., \setupOne{} and \setupSev{}).
Each experiment setup represents an experiment in which the \yasa{} sampling algorithm is executed to either compute a sample that achieves pair-wise coverage (i.e., \setupOne{}) or three-wise coverage (i.e., \setupSev{}).
For our evaluation, we use the implementation of the \yasa{} sampling algorithm that is included in the FeatureIDE library version 3.10.%~\footnote{add link to fide library}.
We execute each experiment on all subject systems specified in \autoref{tab:experimentSetup} ten times, to mitigate random influences on our results. 
During the experiment executions, we measure the time it takes to compute the sample (i.e., sampling time), by setting timestamps before and after executing the \yasa{} sampling algorithm.
We measure the sample size and the achieved pair-wise and three-wise coverage of the resulting samples, by using utility functionality that is included in FeatureIDE library version 3.10.%~\footnote{add link to library}. 

In \autoref{tab:experimentSetup} we specify five experiment setups (i.e., \setupTwo{} to \setupSix{}) each representing an experiment in which the \mulTiWise{} sampling algorithm is executed. 
We execute each experiment setup ten times on each subject system specified in \autoref{tab:subjectsystemTable}.
In each execution, we assign the specified number of features (see \autoref{tab:experimentSetup}) to the respective feature groups by randomly choosing which feature gets assigned to which group. 
By doing so we reduce the experiment bias of assigning the same features to the same t-wise feature interaction groups, while still ensuring that the relative number of features in each group stays the same throughout each experiment execution.
for each experiment execution, we measure the same metrics (i.e., sampling time, sample time, pair-wise coverage, and three-wise coverage) as for executing the \yasa{} sampling algorithm using the same measuring methods.
\subsection{Results}
\label{subsec:results}
\begin{figure}[h!]
\centering
\includegraphics[width=0.7\textwidth]{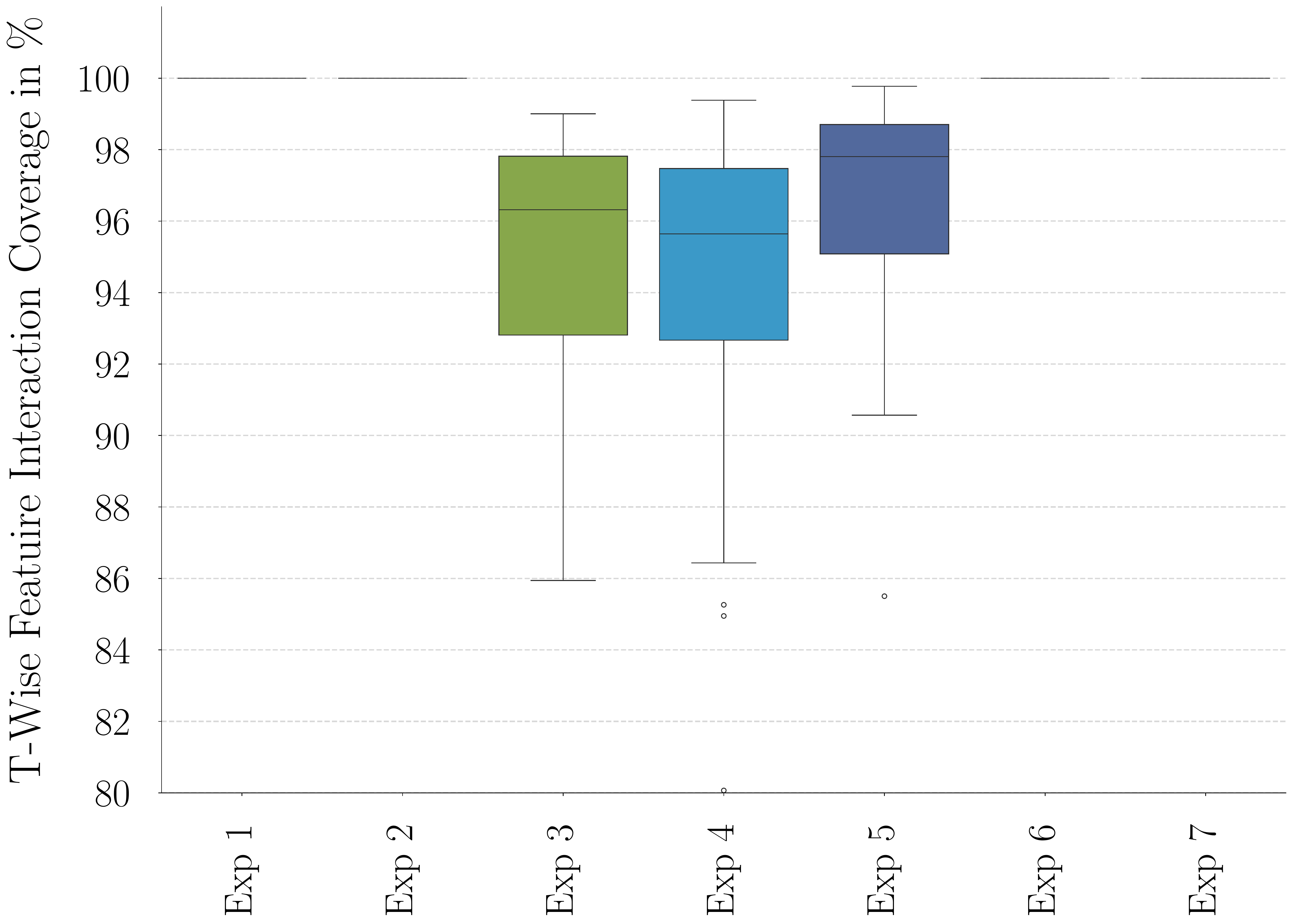}
\caption{Results of measuring pair-wise ($\twise = 2$) coverage.
}
\label{fig:T2_coverage}
\end{figure}

\begin{figure}[h!]
    \centering
    \includegraphics[width=0.7\textwidth]{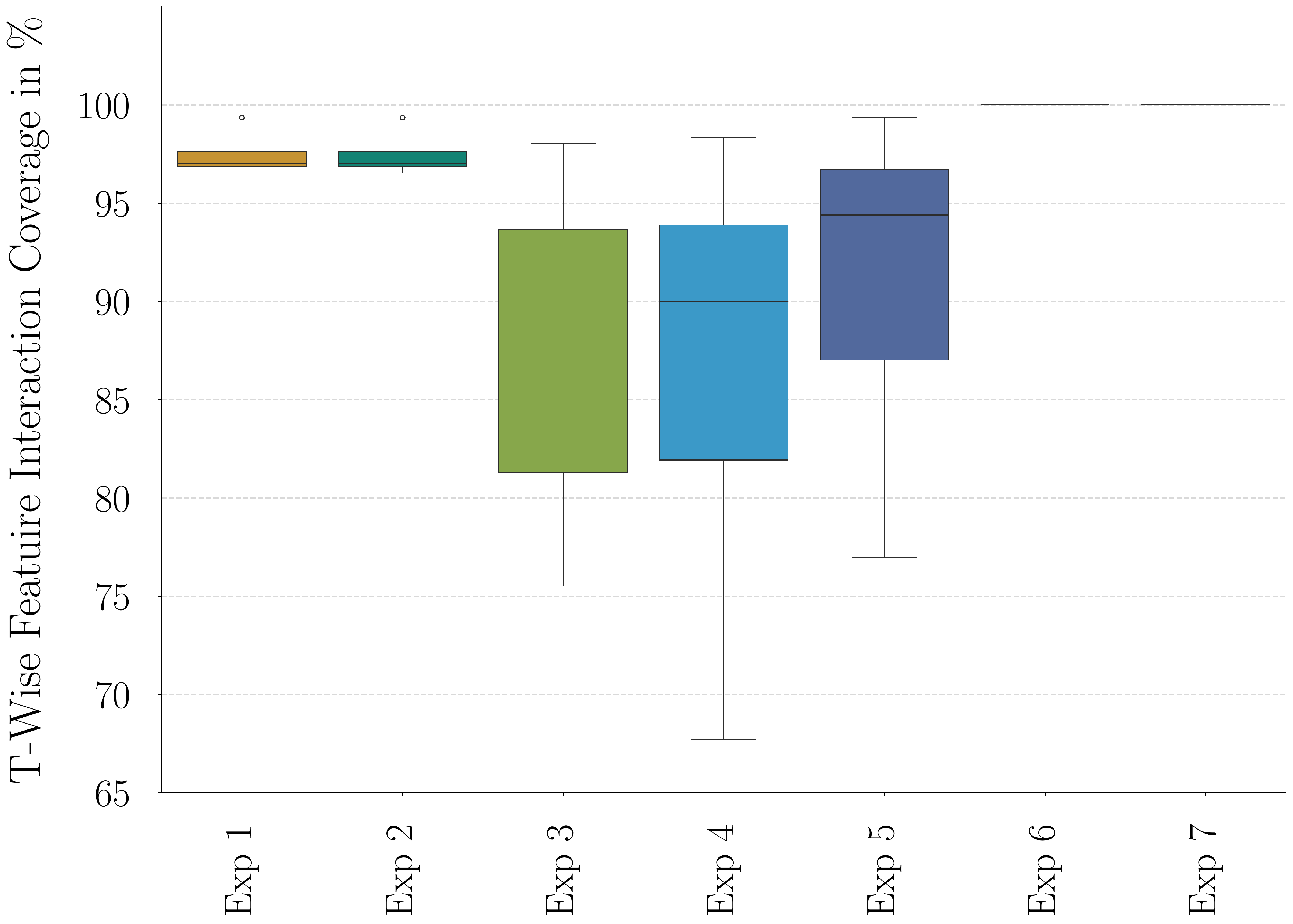}
    \caption{Results of measuring three-wise ($\twise = 3$) coverage.
    }
    \label{fig:T3_coverage}
\end{figure}

\begin{figure}[h!]
    \centering
    \includegraphics[width=0.7\textwidth]{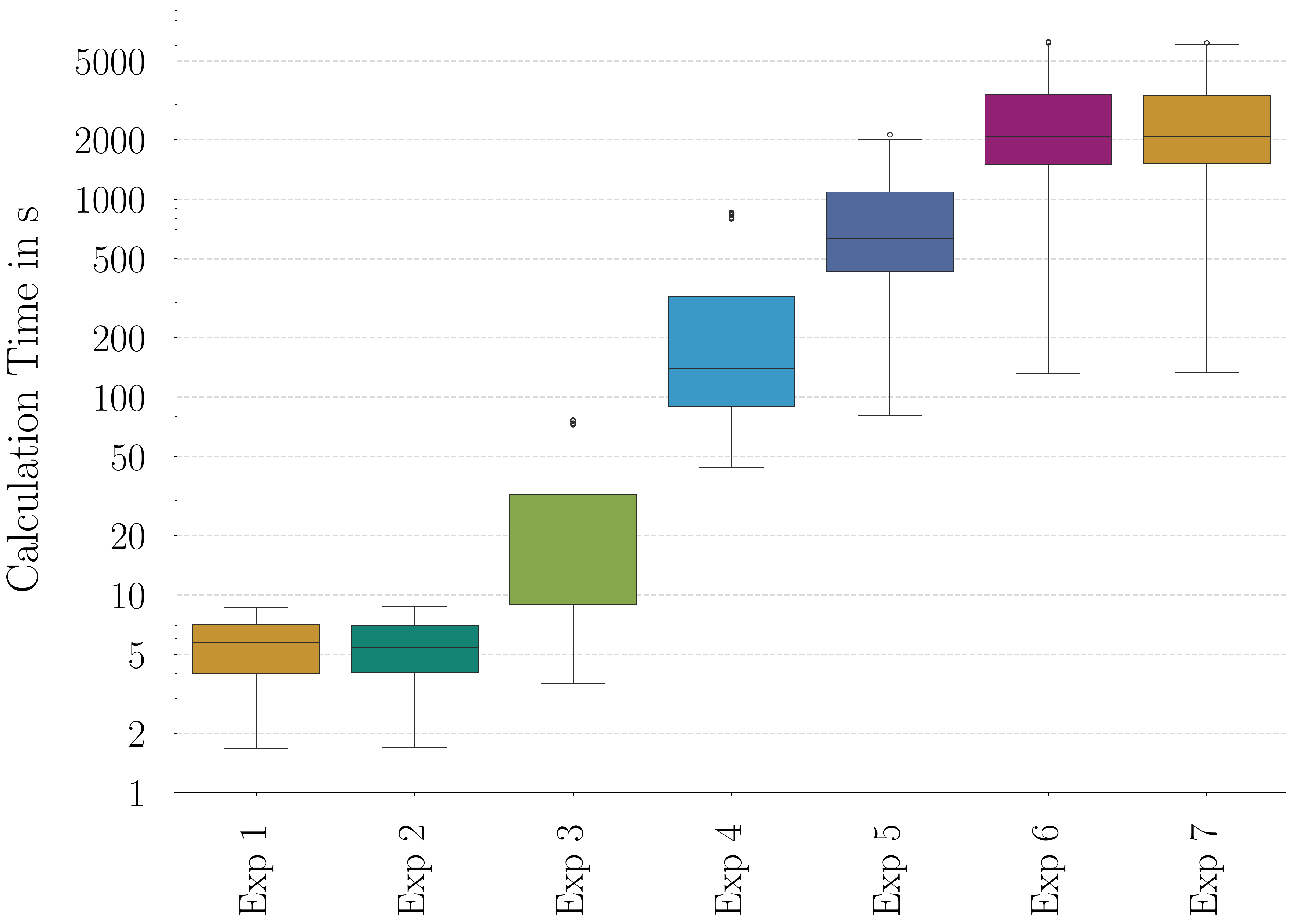}
    \caption{Results of measuring sampling time.
    %for our experiment configurations. %Each box represents results measured for the respective experiment configuration (Label on the X-Axis). The boxes aggregate the results measured for all subject systems.
    }
    \label{fig:time}
\end{figure}

\begin{figure}[h!]
    \centering
    \includegraphics[width=0.7\textwidth]{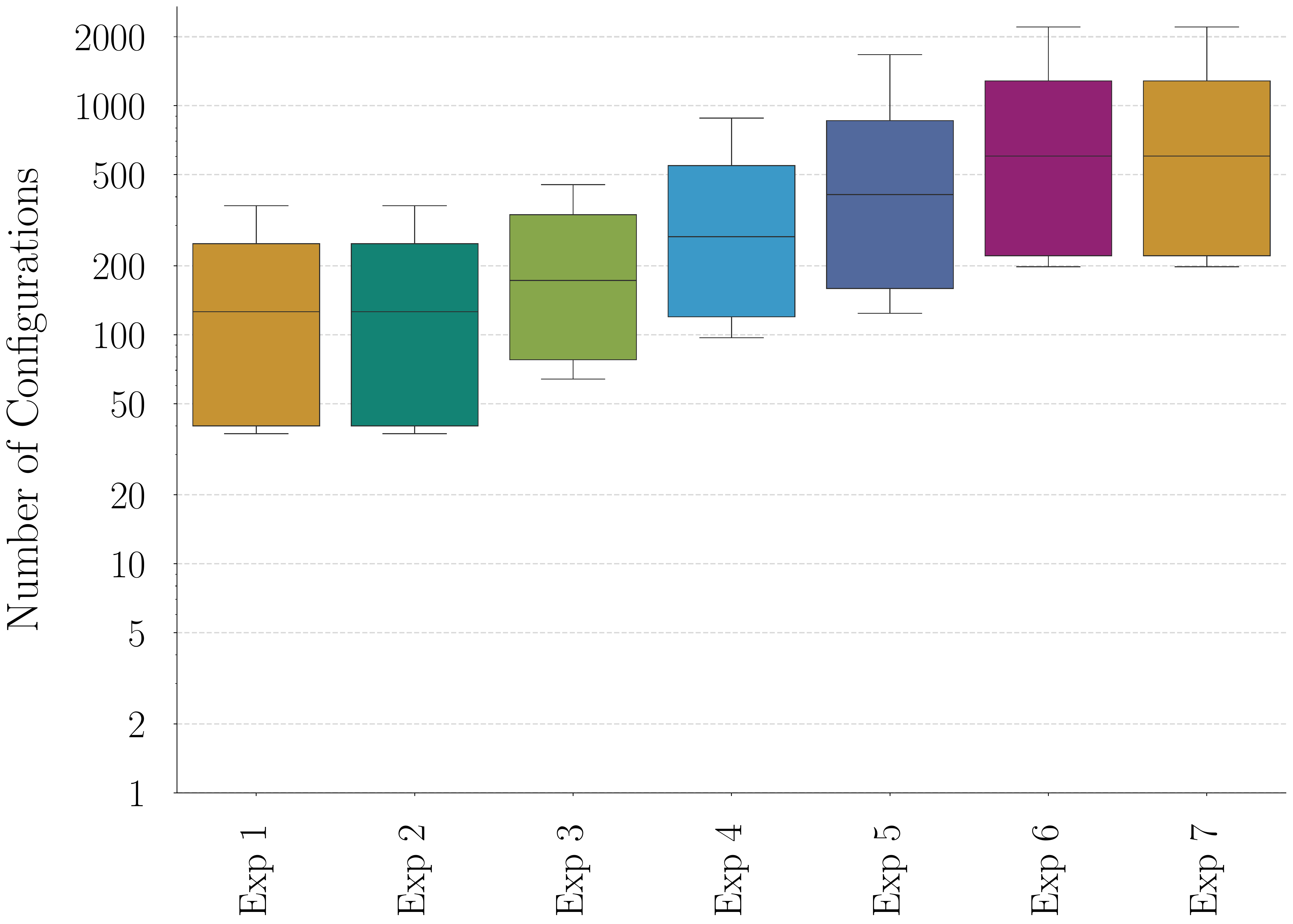}
    \caption{Results of measuring the sample size. 
    }
    \label{fig:size}
\end{figure}

We use a series of boxplots to visualize the results of measuring \emph{Coverage t2} (see \autoref{fig:T2_coverage}), \emph{Coverage t3} (see \autoref{fig:T3_coverage}), \emph{Sample Time} (see \autoref{fig:time}), and \emph{Sample Size} (see \autoref{fig:size}). 
%Each sequence consists of a boxplot for each subject system in our experiments (i.e., \emph{Busybox}, \emph{Fiasco}, \emph{Soletta}, and \emph{UCLibc-NG}). 
Each boxplot accumulates the results measured for all subject systems. 
We show the accumulated results to keep this paper concise and provide a detailed overview of our results in our results package online~\footnote{\footResults} for reference.
The y-axis of each boxplot represents the respective statistic (i.e., \emph{Coverage t2}, \emph{Coverage t3}, \emph{Calculation Time}, and \emph{Sample Size}).
For instance, in \autoref{fig:T2_coverage}, the y-axis represents the pair-wise coverage in percent.
The y-axis for the results of measuring \emph{Coverage t2} and \emph{Coverage t3} are in linear scale, while the y-axis for the results of measuring \emph{Calculation Time} and \emph{Sample Size} are in logarithmic scale. 
We use a logarithmic scale for visualizing \emph{Calculation Time} and \emph{Sample Size} because the results of our experiment setups differ largely for both of these metrics.
Each boxplot contains seven boxes, representing the experiment setups shown in \autoref{tab:experimentSetup}. 
The x-axis of each boxplot shows the identifiers of those setups.
Each boxplot starts with \setupOne{} on the far left, which represents the results for the sampling algorithm \yasa configured to achieve full pair-wise ($\twise = 2$) coverage.
Thereafter, the results for \setupTwo{}, \setupThree{}, \setupFour{}, \setupFive{} \setupSix{} follow.
The last entry on the x-axis always represents results measured for the \yasa sampling algorithm configured to achieve full three-wise ($\twise = 3$) coverage (i.e., \setupSev{}). 
\paragraph{Pair-wise feature interaction coverage}
%%
%%% Figure showing the plots for the coverage t2 analysis
%\input{./004_figures/fig_coverage_t2_accu.tex}
%\input{../004_figures/fig_coverage_t2_accu.tex}
%%

In \autoref{fig:T2_coverage}, we present the results of measuring the pair-wise feature interaction coverage ratio. 
%ratio achieved by the samples from our experiment setups.
The measured median value for our experiments varies between 96\% and 100\%. 
We identify the lowest median coverage (i.e., 96\%) for \setupFour{} and the highest median coverage ratio of 100\% for \setupOne{}, \setupTwo{}, \setupSix{}, and \setupSev{}.
The experiment setup \setupThree{} achieves a higher median value of pair-wie coverage (about 98\%) as \setupFour (). 
The results for \setupFive{} show a higher median value than \setupThree{}.
%
%For the experiment setups \setupOne{}, \setupTwo{}, \setupSix{}, and \setupSev{}, all experiment executions on all subject systems achieve full (i.e., 100\%) pairwise coverage, while the results for experiment setup \setupThree{}, \setupFour{}, and \setupFive{} show spread values. 
%For instance, the results of \setupFour{} show the largest spread with a minimum value of 81\% and a maximum value of 99\%. 
%
\paragraph{Three-wise feature interaction coverage}
%%
%%% Figure showing the plots for the coverage t3 analysis
%\input{../004_figures/fig_coverage_t3.tex}
%\input{./004_figures/fig_coverage_t3_accu.tex}
%%%
In \autoref{fig:T3_coverage}, we present the results of measuring the three-wise feature interaction coverage ratio. 
Overall experiment setups, the median values differ between a maximum of 100\% (achieved by \setupSix{} and \setupSev{}) and a minimum value of 90\% (achieved by \setupThree{} and \setupFour{}). 
The experiment setups \setupOne{} and \setupTwo{} achieve the second highest three-wise coverage ratio with a median value of 96\%. 
\setupFive{} achieves a coverage ratio of 94\% in median.
%For the experiment setups \setupThree{}, \setupFour{}, and \setupFive{}, we see large differences between the results of the experiment executions on our subject systems
%For instance, the lower (about 68\%) and upper (about 98\%) whiskers of \setupFour{} differ by 30\% from each other, and the lower (about 78\%) and upper (about 94\%) quartiles for this experiment setup differ by 16\%. 
%Differences between results for \setupOne{} and \setupTwo{} (about 2\% between lower and upper whiskers) are visibly smaller compared to \setupFour{}. 
%For \setupSix and \setupSev{}, all experiment executions on all subject systems achieve full (i.e., 100\%) three-wise feature interaction coverage. 

\paragraph{Sampling time}
%%
%%% Figure showing the plots for the coverage time analysis
%\input{../004_figures/fig_time.tex}
%\input{./004_figures/fig_time_accu.tex}
In \autoref{fig:time}, we show the time it takes to compute a sample in our experiment setups (i.e., sampling time). 
We measure the shortest median sampling time (about 5 seconds) for \setupOne{} and \setupTwo{}.
The median sampling time grows exponentially from \setupTwo (about 5 seconds) to \setupSix{about 2000 seconds}. 
The exponential growth appears as a linear growth in \autoref{fig:time} because of the logarithmic scale of the y-axis. 
For \setupSev{}, we measure the same median sampling time (about 2000 seconds) as for \setupSix{} 
%The lower quartile group of \setupSev{} is larger than that of the upper quartile group, indicating that the average calculation time is below the median value of 950 seconds.
%On the contrary, the lower quartile group of the experiment setups \setupThree{}, \setupFour{}, \setupFive{}, and \setupSix{} is smaller than the upper quartile group, indicating that the runtime of those setups is in average higher than the median value.
%For \setupOne{} and \setupTwo{}, the lower and upper quartile groups are similarly sized, showing that the average runtime for those experiment setups is close to the median value of 3 seconds.

\paragraph{Sample Size}
%%
%%% Figure showing the plots for the coverage sample size analysis
%\input{../004_figures/fig_size.tex}
%\input{./004_figures/fig_size_accu.tex}
%%%
%%
In \autoref{fig:size}, we show the results of measuring the sample size of the samples resulting from executing our experiment setups on all subject systems.
We measure the smallest sample sizes for the experiment setups \setupOne{} and \setupTwo{}, with a median value of 120.
From \setupTwo{} to \setupSev{}, the sample size grows exponentially from a median value of 120 to a median value of 800. 
In \autoref{fig:size}, the exponential growth in sample size appears linear because of the logarithmic scale of the y-axis. 
\subsection{Discussion}
\label{subsec:discussion}
\paragraph{RQ1: How does the size of feature groups influence sampling metrics of \mulTiWise{}?}
%Answering RQ1 requires discussing the results of measuring the statistics for calculation time, sample size, and achieved t-wise coverage (e.g., pair-wise and three-wise). 
%The results of the experiment setups \setupTwo{}, \setupThree{}, \setupFour{}, \setupFive{}, and \setupSix{} are of particular interest for answering this research question because, throughout those experiment setups, we gradually assign more features (i.e., 25\% in each step) from a pair-wise feature interaction group to a three-wise feature interaction group.
%Therefore, the results of these experiment setups reveal the effects of assigning more features to t-wise feature interaction groups with larger values of $\twise$. 

Our results show exponential growth for sample time (see \autoref{fig:time}) and sample size (see \autoref{fig:size}) when more features of the feature model are assigned to t-wise feature interaction groups with higher \twiseText values. 
%Assigning more features to t-wise feature interaction groups with higher values of \twiseText requires covering more t-wise feature interaction tuples. 
%In general, generating a sample that achieves high t-wise feature interaction coverage requires long calculation times, and the resulting sample is typically large. 
Those results align with the theoretical assumption and general observation that achieving t-wise feature interaction coverage for higher values of \twiseText requires more configurations and longer computation times because more t-wise feature interaction tuples must be covered.
Therefore, our results meet our expectations.
%for measuring sampling time and sample size meet our expectations. 

Our results of measuring the achieved pair-wise feature interaction coverage reveal that assigning 100\% of features to a pair-wise feature interaction group (i.e., \setupTwo) and assigning 100\% of features to a three-wise feature interaction group (i.e., \setupThree{}) achieves 100\% pair-wise feature interaction coverage. 
The results confirm our expectation that a sample that covers 100\% of three-wise feature interactions always covers 100\% of pair-wise feature interactions. 
We observe in our results that shifting 25\% of features from the pair-wise feature interaction group to the three-wise feature interaction group (i.e., \setupTwo{}) reduces the achieved pair-wise feature interaction coverage. 
We explain this observation by how \mulTiWise{} covers t-wise feature interactions. 
Considering the two t-wise feature interaction groups of \setupTwo{} as an example, \mulTiWise{} generates all valid pair-wise feature interaction tuples for the features contained in $TG_{t2}$ (i.e., consisting of 75\% of features) and all three-wise feature interaction tuples for the features contained in $TG_{t3}$ (i.e., consisting of 25\% of features), and covers them in the resulting sample. 
However, \mulTiWise{} does not actively generate and cover valid feature interaction tuples between both groups (i.e., feature tuples for 25\% of features).
Hence, we observe a reduced pair-wise feature interaction coverage when comparing \setupTwo{} and \setupThree{}. 
Observing the achieved pair-wise feature interaction coverage for \setupFour{} and \setupFive{} further supports our explanation. 
In \setupFour{}, we shift another set 25\% of features from the pair-wise interaction group $TG_{t2}$ to the three-wise feature interaction group $TG_t3$, leading to 50\% features that are not actively considered by \mulTiWise{}.
Therefore, we expect a reduced pair-wise feature interaction coverage compared to \setupThree, which is visible in our results (see \autoref{fig:T2_coverage}). 
In \setupFive{}, we reduce the number of features contained in $TG_{t2}$ by another 25\% and respectively increase the number of features contained in $TG_{t3}$ by this amount, leading to a smaller set (i.e., 25\%) of not actively considered features. 
According to our expectation, the achieved pair-wise feature interaction coverage of the samples for \setupFive should increase compared to those of \setupFour, which is visible in our results.

Our results of measuring the ratio of achieved three-wise feature interaction coverage (see \autoref{fig:T3_coverage}) reveal that \mulTiWise{} computes samples that achieve 100\% three-wise coverage when 100\% of all features are assigned to a three-wise feature interaction group (i.e., \setupSix{}). 
We expected those results because \mulTiWise{} internally uses the \yasa sampling algorithm, which reliably computes samples that achieve full t-wise feature interaction coverage for given t-values.
%Noteworthy about our results is that assigning 100\% of features to a pair-wise feature interaction group (i.e., \setupTwo{}) results in samples that achieve about 97\% of three-wise feature interaction coverage. 
%We argue that the configurations required to achieve pair-wise feature interaction coverage for our selection of subject systems also cover many three-wise feature interaction tuples. 
%It is known from the literature~\cite{} that many t-wise feature interaction tuples for higher values of $\twise$ can be covered with a small set of configurations. 
%However, finding configurations that cover the last five percent of t-wise feature interaction tuples is difficult and expensive~\cite{}.
%Therefore, our results are in line with the results of the current literature.
The results for our experiment setups \setupTwo{}, \setupThree{}, \setupFour{}, and \setupFive{} show the same pattern that we have observed in the results of measuring pair-wise feature interaction coverage. 
%Splitting the total number of features between $TG_{t2}$ and $TG_{t3}$ reduces the achieved three-wise feature interaction coverage so that the lowest median value is observed for \setupFour{}. 
To explain this observation, we use the same reasoning as for measuring pair-wise feature interaction coverage. 
%Splitting the total number of features between both groups causes \mulTiWise{} to actively compute and cover less t-wise feature interaction tuples, reducing the total coverage value.

We answer RQ1 based on our discussions as follows: 
Changing the percentual assignment of features to feature groups with different values of \twiseText influences the achieved coverage, sample time, and the resulting sample size. 
Assigning more features to t-wise feature interaction groups with higher values of \twiseText increases the sample size and the sampling time exponentially. 
\mulTiWise{} assures that all valid t-wise feature interaction tuples of the different t-wise interaction groups are covered in the resulting sample because it uses \yasa{} sampling algorithm to calculate the intermediate samples for the t-wise feature interaction groups.
However, splitting the total number of features between two feature interaction groups reduces the number of feature interaction tuples that are actively considered by \mulTiWise{}, reducing the overall achieved t-wise feature interaction coverage. 
%
%We conclude that \mulTiWise{} is suitable for covering small sets of features with high t-wise feature interaction coverage by keeping the calculation time and sample size reasonably small. 
%However, in practice, the tradeoff between covering prioritized sets of features with high t-wise feature interaction coverage and reducing overall t-wise feature interaction coverage must be considered.

\paragraph{RQ2: How does \mulTiWise{} compare to state-of-the-art t-wise sampling algorithms?}
%Answering RQ2 requires the comparison of \mulTiWise{} and the state-of-the-art sampling algorithm YASA. 
%The YASA sampling algorithm is commonly used to compute a sample that achieves 100\% feature interaction coverage for a certain value of \twiseText, while \mulTiWise{} is designed to cover defined sets of features with a certain \twiseText value.
%We compare the YASA sampling setups \setupOne (i.e., YASA pair-wise sampling) and \setupSev (i.e., YASA three-wise sampling) with \setupTwo{} and \setupSix{}, which assign 100\% of features to either the pair-wise feature interaction or the three-wise feature interaction group. 
%
Our results of measuring pair-wise and three-wise feature interaction coverage show that \setupOne{} and \setupTwo{} achieve the same pair-wise and three-wise coverage for our subject systems. 
We observe the same for the experiment setups \setupSix{} and \setupSev{}. 
Therefore, we assume that our application of the \yasa{} sampling algorithm for \mulTiWise{} is equivalent to the original algorithm with respect to computing samples that achieve t-wise feature interaction coverage.
Measuring the size of the resulting samples reveals that \setupOne{} (i.e., \yasa{}) and \setupTwo{} (i.e.,\mulTiWise{}) need the same number of configurations to achieve pair-wise coverage.
We observe the same for achieving three-wise feature interaction coverage when comparing the sample size for \setupSix{} and \setupSev{}.
Those results meet our expectations and show that \mulTiWise{} achieves comparable results to the \yasa{} sampling algorithm with respect to sample size.
%
%We observe that \setupSev{} (i.e., YASA three-wise) computes slightly larger samples compared to \setupSix{}, even though both algorithms achieve full pair-wise and three-wise feature interaction coverage.
%The observation contradicts our expectations and results for \setupOne and \setupTwo, which estimate equal results for \setupSix and \setupSev{}. 
%
The results of measuring sampling time show that \setupOne{} (i.e., \yasa{}) and \setupTwo{} (i.e., \mulTiWise{}) need the same time to compute samples that achieve pair-wise feature interaction coverage. 
We see the same when comparing the sampling time of \yasa{} (i.e., \setupSev{}) and \mulTiWise{} (i.e., \setupSix{}) for computing a sample that achieves three-wise feature interaction coverage. 
%The difference in sample time is even more visible when comparing \setupSix{} and \setupSev{}. 
We expected the sampling time for \mulTiWise{} sampling to be higher than the sampling time of \yasa{} because \mulTiWise{} performs preprocessing operations (e.g., assignment of features to feature interaction groups) on all features of the feature model before calling the \yasa sampling algorithm to compute samples.
We argue that the effect of this pre-processing is not visible for our subject systems because they are small enough so that the pre-processing does not influence the sampling time. 
For larger feature models we expect that the sampling time of \mulTiWise{} is slightly larger than that of the \yasa{} sampling algorithm.
%\mulTiWise{} performs preprocessing operations (e.g., assignment of features to feature interaction groups) on all features of the feature model before calling the \yasa sampling algorithm to compute samples. 
%Therefore, we expected that \mulTiWise{} sampling requires more time to compute samples compared to the native \yasa sampling algorithm. 

We answer RQ2 based on our discussion as follows: 
Compared to the state-of-the-art sampling algorithm YASA, \mulTiWise{} achieves equal results with respect to t-wise feature interaction coverage and sample size. 
Our results even reveal that \mulTiWise{} and \yasa{} have comparable sampling times. 
However, for large subject systems, we expect a longer sampling time of \mulTiWise{} because of its pre-processing operations.
%However, the calculation time of \mulTiWise{} can be between 200 seconds to 1000 seconds longer than the calculation time of the \yasa sampling algorithm. 
%We expected higher calculation times for \mulTiWise{} sampling because it assigns features to feature interaction groups.
%The difference in sample time gets more relevant, for high values of \twiseText.
We conclude that \mulTiWise{} may be used as an alternative to \yasa{} when all features of the feature model will be in the same t-wise feature interaction group. 
However, for larger subject systems an overhead in sampling time is to be expected.
%However, the overhead in calculation times indicates using the \yasa{} sampling algorithm when calculation times are critical.

%
\subsection{Threats to Validity}
\label{subsec:threats}
\paragraph{Internal Validity}
A threat to internal validity is based on our choice of preprocessed models for our evaluation. 
The feature models of our subject systems were extracted from kConfig files. 
There are various procedures to transform a kConfig variability model into a feature model, such as using the tseitin transformation~\cite{kuiter2022}. 
Recently, Kuiter et al.~\cite{kuiter2022} discovered that the results of analyzing a feature model of a subject system depend on the transformation of the kConfig file into a feature model file.
We use preprocessed feature models in our experiments, which were used to show the validity of different feature model analysis procedures~\cite{PTR+:SPLC19, PKR+:VAMOS21, PHK+:SPLC23}.
Therefore, we argue that our results align with existing research results, even if the feature models of our subject systems are influenced by the transformation from kConfig models.
Another threat to internal validity is that our implementation uses the FeatureIDE library\footnote{\fideLink} to perform basic feature model analysis tasks.
For instance, we use the implementation of the \yasa sampling algorithm as the base for our \mulTiWise{}.  
%Even though we intensively verified our implementation on smaller feature models, we cannot assure that our results are not biased by how certain analysis methods are implemented in the FeatureIDE library. 
%However, FeatureIDE is widely used in the product-line community, and many existing research tools use the FeatureIDE library as the basis for their experiments~\cite{PHK+:SPLC23, Kri+:SPLC20, HST+:MODELS22}. 
FeatureIDE is widely used in the product-line community, and many existing research tools use the FeatureIDE library as the basis for their experiments~\cite{PHK+:SPLC23, Kri+:SPLC20, HST+:MODELS22}.
Therefore, we argue that the product-line community accepts the correctness of the FeatureIDE library.
% and that even if the use of the library introduces a bias to our results, this bias aligns with existing work.
%% tool implementation
%% tseitin transformation

\paragraph{External Validity}
%% random feature distribution
A threat to external validity is that we assign features randomly to one of the t-wise feature interaction groups of our experiment setups for our evaluation. 
This random assignment may influence the measured sample time and sample size. 
%For instance, if many core features of a feature model are assigned to a t-wise feature interaction group with a high \twiseText value, then we expect a smaller sample size and a shorter sample time. 
We execute every experiment setup of our evaluation ten times to mitigate the resulting bias. 
%We consider the results of each experiment execution in the visualization and discussion of our results. 
%Using boxplots as result visualization enables us to identify extreme outliers caused by the random feature assignment.
%
%% limited use of subject systems
Another threat to external validity is that we consider only five subject systems in our evaluation. 
We cannot assure that performing our evaluation on other subject systems will reveal the same results. 
However, the product-line community widely uses the subject systems in our evaluation to evaluate new concepts~\cite{PTR+:SPLC19, PHK+:SPLC23, KTS+:VaMoS20}.
Hence, our results align with existing work, even if the choice of subject systems introduces a bias.
%We argue that if the choice of subject systems biases our results, they still align with existing work. 

%% only pairwise and three-wise // limited experiment setups
%The third threat to external validity is that we perform our evaluation only with a limited number of experiment setups. 
%For example, smaller increments of assigning features to our pair-wise and three-wise feature interaction groups would allow a more detailed analysis of how changing the feature assignment influences coverage ratio, calculation time, and sample size.

\label{sec:eval}
%
% Related Work
\section{Related Work}
%\textcolor{kitPurple5}{%% begins text color
Existing literature presents many approaches to efficiently generate samples that achieve high t-wise feature interaction coverage~\cite{KKL+:Book13, CM+IST14, LFRE:IWCT15, MK+:ICSE16, VAT+:SPLC18} including sampling algorithms that generate samples that always achieve full t-wise feature interaction coverage~\cite{C:MOR79, CDS+:ISSTA06, CDS+:ISSTA07,  PS+:ICST10, JHF:MODELS11, MHF:SPLC12, AKT+:GPCE16,  KTS+:VaMoS20, Kri+:SPLC20} and algorithms that do not guarantee full t-wise feature interaction coverage~\cite{OGB+:SPLC19, BLM+:FSE20, LS+:ESEC21, BaL+:SPLC22, HF+:EMSE22, HS+:VaMoS24}.
%Studying the literature, we identify sampling algorithms that generate samples that always achieve full t-wise feature interaction coverage~\cite{C:MOR79, CDS+:ISSTA06, CDS+:ISSTA07,  PS+:ICST10, JHF:MODELS11, MHF:SPLC12, AKT+:GPCE16,  KTS+:VaMoS20, Kri+:SPLC20} and algorithms that do not guarantee full t-wise feature interaction coverage of the generated samples~\cite{OGB+:SPLC19, BLM+:FSE20, LS+:ESEC21, BaL+:SPLC22, HF+:EMSE22, HS+:VaMoS24}.
In the following, we discuss the relation of \mulTiWise{} to sampling algorithms from both categories.
%In the following section, we present an overview of sampling algorithms from both categories and discuss how our \mulTiWise{} differs from existing approaches.  
%} %% ends text color
%The first strategy is to cover all t-wise feature interaction tuples of the configurable system in at least one of the resulting configurations (i.e., t-wise feature interaction sampling)~\cite{C:MOR79, CDS+:ISSTA06, CDS+:ISSTA07,  PS+:ICST10, JHF:MODELS11, MHF:SPLC12, AKT+:GPCE16,  KTS+:VaMoS20, Kri+:SPLC20}.
%The second strategy is to replicate the overall configuration space as truthfully as possible (i.e., uniform random sampling)~\cite{OGB+:SPLC19, BLM+:FSE20, LS+:ESEC21, BaL+:SPLC22, HF+:EMSE22, HS+:VaMoS24}.  
%
%\textcolor{kitPurple5}{%% begins text color
\paragraph*{Full T-Wise Feature Interaction Coverage}
T-wise feature interaction sampling approaches aim to cover each valid t-wise feature interaction tuple of the configurable system in at least one configuration of the resulting sample while keeping the number of configurations as small as possible~\cite{VAT+:SPLC18}.
Cohen et al.~\cite{CDS+:ISSTA06} show that finding a minimal set of configurations that covers all valid t-wise feature interaction tuples of a configurable system is an instance of the set-covering problem. 
Chvatal's Algorithm~\cite{C:MOR79} is a procedure to solve the general set-covering problem using a greedy heuristic but building and verifying all possible solutions, as required by Chvatal's Algorithm~\cite{C:MOR79} is not feasible in practice.  
Johansen et al.~\cite{JHF:MODELS11} show that Boolean satisfiability solvers (SAT-solvers) make generating samples with Chvatal's algorithm possible for real-world configurable systems by providing their sampling algorithm ICPL~\cite{MHF:SPLC12}.
%They also reduce the runtime of generating sets of configurations by providing their sampling algorithm ICPL~\cite{MHF:SPLC12}.
ICPL introduces removing core and dead features, removing invalid t-wise feature interaction tuples, and parallelization to generate samples faster compared to Chvatal's algorithms. 
Over the past decades, many authors proposed various tweaks to adapt Chvatal's algorithm to reduce the runtime or sample size~\cite{MK+:ICSE16, VAT+:SPLC18}. 
Currently, the \yasa{} sampling algorithm proposed by Krieter et al.~\cite{KTS+:VaMoS20, Kri+:SPLC20} is the best-performing algorithm with respect to sample size and sampling time.
In addition, \yasa{} provides functionalities such as generating samples that achieve t-wise feature interaction coverage for only a subset of features. 
%} %% ends text color
%To realize our sampling procedure, we use YASA~\cite{KTS+:VaMoS20, Kri+:SPLC20} as our core-covering algorithm and utilize its functionality to consider previous samples and to generate samples that achieve t-wise feature interaction coverage for subsets of features. 
%
%\textcolor{kitPurple5}{%% begins text color
\paragraph*{Close to Full T-Wise Feature Interaction Coverage} 
Recently, many sampling algorithms were proposed that trade full t-wise feature interaction coverage against reduced algorithm runtime and comparable sample sizes~\cite{HS+:VaMoS24}.  
Oh et al.~\cite{OGB+:SPLC19} present a procedure to generate samples that achieves partial but not full t-wise feature interaction coverage. 
Their procedure uses a \#SAT-solver to compute the number of valid configurations of the configurable system, which makes uniform random sampling in a tractable time possible. 
The evaluation of Oh et al.~\cite{OGB+:SPLC19} shows that their uniform random sampling approach does not achieve full t-wise feature interaction coverage. 
Recent work~\cite{BLM+:FSE20, LS+:ESEC21, BaL+:SPLC22, HS+:VaMoS24} identifies that t-wise feature interaction tuples are not uniformly distributed throughout the configuration space because of constraints in the feature model.
Therefore, uniform random sampling approaches need guidance to find uncovered t-wise feature interaction tuples with fewer configurations.
Baranov et al.~\cite{BLM+:FSE20, BaL+:SPLC22} propose an adaptive weighted sampling algorithm \textsc{Baital}, which achieves high t-wise feature interaction coverage values.
In contrast to typical uniform random sampling algorithms, \textsc{Baital} incrementally adapts the weights of selectable literals based on previously selected configurations, thus counteracting the non-uniform distribution of t-wise feature interaction tuples.
%By adapting the weights of the selected literals, \textsc{Baital} counteracts the non-uniform distribution of t-wise feature interaction tupels.
Lou et al.~\cite{LS+:ESEC21} present LS-Sampling, a local search-based sampling approach that achieves high t-wise feature interaction coverage values. 
LS-Sampling iteratively selects configurations from the configuration space using a local search framework and updates the internal probability of selecting certain feature tuples based on the previously chosen configuration so that t-wise feature interaction coverage of the resulting sample can be maximized faster. 
%After choosing a configuration LS-Sampling, updates the internal probability of selecting certain feature tuples based on the previously chosen configuration so that the t-wise feature interaction coverage of the resulting sample can be maximized as fast as possible. 
%} %% ends text color
%
%\textcolor{kitPurple5}{%% begins text color
\paragraph*{Advancement Beyond State-Of-the-Art}
None of the existing sampling algorithms considers covering subsets of features with different t-wise feature interaction coverage values to reduce the resulting sample size. 
\mulTiWise{} fills this gap in the existing research, providing the possibility to sample subsets of features from a configurable system with different t-values. 
We are also the first to show that the sample size can be reduced by covering only a small subset of features with high t-values while trading off full t-wise feature interaction coverage. 
%} %% ends text color
\label{sec:relatedWork}
%
% Conclusion
\section{Conclusion and Future Work}
In this paper, we question the necessity of achieving full t-wise feature interaction coverage for all features equally and present \mulTiWise{} as a novel sampling approach that enables a trade-off between equally achieving full t-wise feature interaction coverage for all features and achieving full t-wise feature interaction coverage for specified feature groups.
In practice, sample-based testing for highly configurable systems strives to save testing resources by reducing the number of configurations for testing. 
However, achieving full t-wise feature interaction coverage for higher values of \twiseText still requires many configurations because of the enormous number of t-wise feature interaction tuples that must be covered. 
As a solution idea, we propose to reduce the number of feature interactions to be covered by defining groups of features for which high t-wise feature interaction coverage will be achieved and groups of features for which t-wise feature interaction coverage is neglectable.  
We present \mulTiWise{} as a sampling approach that considers those feature groups during sample generation, to enable a trade-off between a large sample that achieves full t-wise feature interaction coverage for all features and smaller samples that do so only for certain groups of features.
%With \mulTiWise{}, we present (to the best of our knowledge) the first sampling approach that reduces the number of configurations for testing while assuring that groups of critical features are covered with high ($\twise >= 3$) \twiseText values.
%As part of our sampling approach, we present the concept of assigning features of the feature model to different t-wise feature interaction groups based on their criticality.  
%A t-wise feature interaction group is characterized by the assigned features and a \twiseText value, so each feature group may have different \twiseText values. 
%\mulTiWise sampling calculates a sample that coverers all valid t-wise feature interaction for each group critical features individually, trading away uniform t-wise feature interaction coverage.
We evaluate our approach on four subject systems from real-world applications (i.e., \busybox, \fiasco, \soletta, \uclibc).
Our results show that \mulTiWise{} reduces the calculation time and sample size when only a small number of features are assigned to critical feature groups with high \twiseText values. 
As a tradeoff for reduced sample size and sampling time, \mulTiWise{} does not achieve full coverage for all features equally.
The performance of \mulTiWise{} is comparable to the state-of-the-art sampling algorithm \yasa{}.
Hence, we argue that \mulTiWise{} is an alternative to established sampling approaches if the criticality of features is known.

As future work, we aim to evaluate \mulTiWise{} on more feature models typically used by the community\footnote{\footAddSubjectSystem} to show the feasibility of our approach on a broader set of feature models.
%We will connect with existing research~\cite{PES+:MASE20} to identify metrics to identify critical features and categorize them into t-wise feature interaction groups in practice. 
We expect high synergies between solution-space sampling~\cite{HPS+:GPCE22} and \mulTiWise{} because critical feature groups can be directly identified from attributes of implementation artefacts.
Therefore, we aim to connect both research areas in the future.
%Furthermore, we will investigate if \mulTiWise{} is applicable in \emph{Generaic Solution Space Sampling for Multi-domain Product Lines}. 

%----------------------------------------------------------------------------
%%
%\begin{acks}
%	\input{./002_sections/acks.tex}
%\end{acks}
%\newpage
%------------------------------------------------------------------------

\bibliographystyle{unsrtnat}
\bibliography{./003_bib/multiwise_sampling}  %%% Uncomment this line and comment out the ``thebibliography'' section below to use the external .bib file (using bibtex) .

\begin{thebibliography}{50}
\providecommand{\natexlab}[1]{#1}
\providecommand{\url}[1]{\texttt{#1}}
\expandafter\ifx\csname urlstyle\endcsname\relax
  \providecommand{\doi}[1]{doi: #1}\else
  \providecommand{\doi}{doi: \begingroup \urlstyle{rm}\Url}\fi

\bibitem[Ammann and Offutt(2016)]{AmO+:16}
Paul Ammann and Jeff Offutt.
\newblock \emph{Introduction to software testing}.
\newblock Cambridge University Press, 2016.

\bibitem[McGregor(2010)]{M10}
John McGregor.
\newblock {Testing a Software Product Line}.
\newblock In \emph{Testing Techniques in Software Engineering}, pages 104--140.
  2010.

\bibitem[Duvall et~al.(2007)Duvall, Matyas, and Glover]{DMG:07}
Paul~M Duvall, Steve Matyas, and Andrew Glover.
\newblock \emph{Continuous integration: improving software quality and reducing
  risk}.
\newblock Pearson Education, 2007.

\bibitem[Fowler and Foemmel(2006)]{FF:06}
Martin Fowler and Matthew Foemmel.
\newblock Continuous integration, 2006.

\bibitem[Meyer(2014)]{Mey+:IEEESoftware14}
Mathias Meyer.
\newblock Continuous integration and its tools.
\newblock \emph{IEEE Software}, 31\penalty0 (3):\penalty0 14--16, 2014.

\bibitem[Pohl et~al.(2005)Pohl, B\"{o}ckle, and van~der Linden]{PBL05}
Klaus Pohl, G\"{u}nter B\"{o}ckle, and Frank~J. van~der Linden.
\newblock \emph{{Software Product Line Engineering: Foundations, Principles and
  Techniques}}.
\newblock 2005.

\bibitem[Pett et~al.(2019)Pett, Th\"{u}m, Runge, Krieter, Lochau, and
  Schaefer]{PTR+:SPLC19}
Tobias Pett, Thomas Th\"{u}m, Tobias Runge, Sebastian Krieter, Malte Lochau,
  and Ina Schaefer.
\newblock Product {Sampling} for {Product} {Lines}: {The} {Scalability}
  {Challenge}.
\newblock In \emph{Proceedings of the 23rd {International} {Systems} and
  {Software} {Product} {Line} {Conference} - {Volume} {A}}, {SPLC} '19, pages
  78--83, New York, NY, USA, September 2019. Association for Computing
  Machinery.
\newblock ISBN 9781450371384.
\newblock \doi{10.1145/3336294.3336322}.
\newblock URL \url{https://dl.acm.org/doi/10.1145/3336294.3336322}.

\bibitem[Apel et~al.(2013)Apel, Batory, K\"astner, and Saake]{ABKS13}
Sven Apel, Don Batory, Christian K\"astner, and Gunter Saake.
\newblock \emph{{Feature-Oriented Software Product Lines}}.
\newblock 2013.

\bibitem[Engstr\"{o}m and Runeson(2011)]{ER:IST11}
Emelie Engstr\"{o}m and Per Runeson.
\newblock {Software Product Line Testing - A Systematic Mapping Study}.
\newblock 53:\penalty0 2--13, January 2011.
\newblock ISSN 0950-5849.
\newblock \doi{http://dx.doi.org/10.1016/j.infsof.2010.05.011}.

\bibitem[Lee et~al.(2012)Lee, Kang, and Lee]{LKL:SPLC12}
Jihyun Lee, Sungwon Kang, and Danhyung Lee.
\newblock {A Survey on Software Product Line Testing}.
\newblock pages 31--40, 2012.

\bibitem[Halin et~al.(2019)Halin, Nuttinck, Acher, Devroey, Perrouin, and
  Baudry]{HNA+:EMSE19}
Axel Halin, Alexandre Nuttinck, Mathieu Acher, Xavier Devroey, Gilles Perrouin,
  and Benoit Baudry.
\newblock {Test them all, is it worth it? Assessing configuration sampling on
  the JHipster Web development stack}.
\newblock \emph{Empirical Software Engineering}, 24\penalty0 (2):\penalty0
  674--717, April 2019.
\newblock ISSN 1573-7616.
\newblock \doi{10.1007/s10664-018-9635-4}.
\newblock URL \url{https://doi.org/10.1007/s10664-018-9635-4}.

\bibitem[Medeiros et~al.(2015)Medeiros, K{\"a}stner, Ribeiro, Nadi, and
  Gheyi]{MKR+:ECOOP15}
Fl{\'a}vio Medeiros, Christian K{\"a}stner, M{\'a}rcio Ribeiro, Sarah Nadi, and
  Rohit Gheyi.
\newblock {The Love/Hate Relationship with the C Preprocessor: An Interview
  Study}.
\newblock In \emph{29th European Conference on Object-Oriented Programming
  (ECOOP 2015)}, pages 495--518. Schloss Dagstuhl--Leibniz-Zentrum fuer
  Informatik, 2015.

\bibitem[Varshosaz et~al.(2018)Varshosaz, Al-Hajjaji, Th\"um, Runge, Mousavi,
  and Schaefer]{VAT+:SPLC18}
Mahsa Varshosaz, Mustafa Al-Hajjaji, Thomas Th\"um, Tobias Runge, Mohammad~Reza
  Mousavi, and Ina Schaefer.
\newblock {A Classification of Product Sampling for Software Product Lines}.
\newblock pages 1--13, 2018.

\bibitem[Cohen et~al.(2008)Cohen, Dwyer, and Shi]{CDS:TSE08}
Myra~B. Cohen, Matthew~B. Dwyer, and Jiangfan Shi.
\newblock {Constructing Interaction Test Suites for Highly-Configurable Systems
  in the Presence of Constraints: A Greedy Approach}.
\newblock 34\penalty0 (5):\penalty0 633--650, 2008.

\bibitem[Marijan et~al.(2013)Marijan, Gotlieb, Sen, and Hervieu]{MGSH:SPLC13}
Dusica Marijan, Arnaud Gotlieb, Sagar Sen, and Aymeric Hervieu.
\newblock {Practical Pairwise Testing for Software Product Lines}.
\newblock pages 227--235, 2013.

\bibitem[Krieter et~al.(2020)Krieter, Th\"{u}m, Schulze, Saake, and
  Leich]{KTS+:VaMoS20}
Sebastian Krieter, Thomas Th\"{u}m, Sandro Schulze, Gunter Saake, and Thomas
  Leich.
\newblock {YASA: Yet Another Sampling Algorithm}.
\newblock 2020.

\bibitem[Oh et~al.(2017)Oh, Batory, Myers, and Siegmund]{OBM+:FSE17}
Jeho Oh, Don Batory, Margaret Myers, and Norbert Siegmund.
\newblock {Finding Near-Optimal Configurations in Product Lines by Random
  Sampling}.
\newblock pages 61--71, August 2017.
\newblock \doi{10.1145/3106237.3106273}.

\bibitem[Munoz et~al.(2019)Munoz, Oh, Pinto, Fuentes, and Batory]{MOP+:SPLC19}
Daniel-Jesus Munoz, Jeho Oh, M\'{o}nica Pinto, Lidia Fuentes, and Don Batory.
\newblock {Uniform Random Sampling Product Configurations of Feature Models
  That Have Numerical Features}.
\newblock pages 289--301, 2019.

\bibitem[Luo et~al.(2023)Luo, Song, Zhao, Li, Cai, and Hu]{LSZ+:SPLC23}
Chuan Luo, Jianping Song, Qiyuan Zhao, Yibei Li, Shaowei Cai, and Chunming Hu.
\newblock Generating pairwise covering arrays for highly configurable software
  systems.
\newblock In \emph{Proceedings of the 27th {ACM} {International} {Systems} and
  {Software} {Product} {Line} {Conference} - {Volume} {A}}, volume~A of
  \emph{SPLC '23}, page 261–267, New York, NY, USA, August 2023. Association
  for Computing Machinery.
\newblock ISBN 9798400700910.
\newblock \doi{10.1145/3579027.3608998}.
\newblock URL \url{https://doi.org/10.1145/3579027.3608998}.

\bibitem[Bombarda et~al.(2023)Bombarda, Bonfanti, and Gargantini]{BBG+:SPLC23}
Andrea Bombarda, Silvia Bonfanti, and Angelo Gargantini.
\newblock On the reuse of existing configurations for testing evolving feature
  models.
\newblock In \emph{Proceedings of the 27th ACM International Systems and
  Software Product Line Conference - Volume B}, SPLC '23, page 67–76, New
  York, NY, USA, 2023. Association for Computing Machinery.
\newblock ISBN 9798400700927.
\newblock \doi{10.1145/3579028.3609017}.
\newblock URL \url{https://doi.org/10.1145/3579028.3609017}.

\bibitem[Pett et~al.(2023)Pett, He\ss{}, Krieter, Th\"{u}m, and
  Schaefer]{PHK+:SPLC23}
Tobias Pett, Tobias He\ss{}, Sebastian Krieter, Thomas Th\"{u}m, and Ina
  Schaefer.
\newblock Continuous {T}-{Wise} {Coverage}.
\newblock In \emph{Proceedings of the 27th ACM International Systems and
  Software Product Line Conference - Volume A}, volume~A of \emph{{SPLC} '23},
  pages 87--98, New York, NY, USA, August 2023. Association for Computing
  Machinery.
\newblock ISBN 9798400700910.
\newblock \doi{10.1145/3579027.3608980}.
\newblock URL \url{https://doi.org/10.1145/3579027.3608980}.

\bibitem[Al-Hajjaji et~al.(2014)Al-Hajjaji, Th\"um, Meinicke, Lochau, and
  Saake]{ATM+:SPLC14subsumedbyATL+:SoSyM19}
Mustafa Al-Hajjaji, Thomas Th\"um, Jens Meinicke, Malte Lochau, and Gunter
  Saake.
\newblock {Similarity-Based Prioritization in Software Product-Line Testing}.
\newblock pages 197--206, September 2014.
\newblock ISBN 978-1-4503-2740-4.
\newblock \doi{10.1145/2648511.2648532}.

\bibitem[Pett et~al.(2020)Pett, Eichhorn, and Schaefer]{PES+:MASE20}
Tobias Pett, Domenik Eichhorn, and Ina Schaefer.
\newblock Risk-based compatibility analysis in automotive systems engineering.
\newblock In \emph{Proceedings of the 23rd {ACM}/{IEEE} {International}
  {Conference} on {Model} {Driven} {Engineering} {Languages} and {Systems}:
  {Companion} {Proceedings}}, {MODELS} '20, pages 1--10, New York, NY, USA,
  October 2020. Association for Computing Machinery.
\newblock ISBN 9781450381352.
\newblock \doi{10.1145/3417990.3421263}.
\newblock URL \url{https://dl.acm.org/doi/10.1145/3417990.3421263}.

\bibitem[Lachmann et~al.(2017)Lachmann, Beddig, Lity, Schulze, and
  Schaefer]{LB+:VaMoS17}
Remo Lachmann, Simon Beddig, Sascha Lity, Sandro Schulze, and Ina Schaefer.
\newblock Risk-based integration testing of software product lines.
\newblock In \emph{Proceedings of the 11th {International} {Workshop} on
  {Variability} {Modelling} of {Software}-{Intensive} {Systems}}, {VaMoS} '17,
  pages 52--59, New York, NY, USA, February 2017. Association for Computing
  Machinery.
\newblock ISBN 9781450348119.
\newblock \doi{10.1145/3023956.3023958}.
\newblock URL \url{https://dl.acm.org/doi/10.1145/3023956.3023958}.

\bibitem[Amland(2000)]{Aml+:JSS00}
Ståle Amland.
\newblock Risk-based testing:: {Risk} analysis fundamentals and metrics for
  software testing including a financial application case study.
\newblock \emph{Journal of Systems and Software}, 53\penalty0 (3):\penalty0
  287--295, September 2000.
\newblock ISSN 0164-1212.
\newblock \doi{10.1016/S0164-1212(00)00019-4}.
\newblock URL
  \url{https://www.sciencedirect.com/science/article/pii/S0164121200000194}.

\bibitem[Michalik and Weyns(2011)]{MiW+:2011SAC11}
Bartosz Michalik and Danny Weyns.
\newblock Towards a {Solution} for {Change} {Impact} {Analysis} of {Software}
  {Product} {Line} {Products}.
\newblock In \emph{2011 {Ninth} {Working} {IEEE}/{IFIP} {Conference} on
  {Software} {Architecture}}, pages 290--293, June 2011.
\newblock \doi{10.1109/WICSA.2011.45}.
\newblock URL \url{https://ieeexplore.ieee.org/document/5959729}.

\bibitem[Krieter(2020)]{Kri+:SPLC20}
Sebastian Krieter.
\newblock Large-scale {T}-wise interaction sampling using {YASA}.
\newblock In \emph{Proceedings of the 24th {ACM} {Conference} on {Systems} and
  {Software} {Product} {Line}: {Volume} {A} - {Volume} {A}}, {SPLC} '20, pages
  1--4, New York, NY, USA, October 2020. Association for Computing Machinery.
\newblock ISBN 9781450375696.
\newblock \doi{10.1145/3382025.3414989}.
\newblock URL \url{https://dl.acm.org/doi/10.1145/3382025.3414989}.

\bibitem[Clements and Northrop(2001)]{CN01}
Paul Clements and Linda Northrop.
\newblock \emph{{Software Product Lines: Practices and Patterns}}.
\newblock 2001.

\bibitem[Medeiros et~al.(2016)Medeiros, Kästner, Ribeiro, Gheyi, and
  Apel]{MK+:ICSE16}
Flávio Medeiros, Christian Kästner, Márcio Ribeiro, Rohit Gheyi, and Sven
  Apel.
\newblock A comparison of 10 sampling algorithms for configurable systems.
\newblock In \emph{Proceedings of the 38th {International} {Conference} on
  {Software} {Engineering}}, {ICSE} '16, pages 643--654, New York, NY, USA, May
  2016. Association for Computing Machinery.
\newblock ISBN 9781450339001.
\newblock \doi{10.1145/2884781.2884793}.
\newblock URL \url{https://dl.acm.org/doi/10.1145/2884781.2884793}.

\bibitem[Cohen et~al.(2007)Cohen, Dwyer, and Shi]{CDS+:ISSTA07}
Myra~B. Cohen, Matthew~B. Dwyer, and Jiangfan Shi.
\newblock Interaction testing of highly-configurable systems in the presence of
  constraints.
\newblock In \emph{Proceedings of the 2007 international symposium on
  {Software} testing and analysis}, {ISSTA} '07, pages 129--139, New York, NY,
  USA, July 2007. Association for Computing Machinery.
\newblock ISBN 9781595937346.
\newblock \doi{10.1145/1273463.1273482}.
\newblock URL \url{https://dl.acm.org/doi/10.1145/1273463.1273482}.

\bibitem[Johansen et~al.(2012{\natexlab{a}})Johansen, Haugen, Fleurey,
  Eldegard, and Syversen]{JHF+:MODELS12}
Martin~Fagereng Johansen, {\O}ystein Haugen, Franck Fleurey, Anne~Grete
  Eldegard, and Torbj{\o}rn Syversen.
\newblock {Generating Better Partial Covering Arrays by Modeling Weights on
  Sub-Product Lines}.
\newblock pages 269--284. 2012{\natexlab{a}}.

\bibitem[Kuiter et~al.(2023)Kuiter, Krieter, Sundermann, Th\"{u}m, and
  Saake]{kuiter2022}
Elias Kuiter, Sebastian Krieter, Chico Sundermann, Thomas Th\"{u}m, and Gunter
  Saake.
\newblock Tseitin or not tseitin? the impact of cnf transformations on
  feature-model analyses.
\newblock In \emph{Proceedings of the 37th IEEE/ACM International Conference on
  Automated Software Engineering}, ASE '22, New York, NY, USA, 2023.
  Association for Computing Machinery.
\newblock ISBN 9781450394758.
\newblock \doi{10.1145/3551349.3556938}.
\newblock URL \url{https://doi.org/10.1145/3551349.3556938}.

\bibitem[Pett et~al.(2021)Pett, Krieter, Runge, Th{\"u}m, Lochau, and
  Schaefer]{PKR+:VAMOS21}
Tobias Pett, Sebastian Krieter, Tobias Runge, Thomas Th{\"u}m, Malte Lochau,
  and Ina Schaefer.
\newblock Stability of {Product}-{Line} {Samplingin} {Continuous}
  {Integration}.
\newblock In \emph{Proceedings of the 15th {International} {Working}
  {Conference} on {Variability} {Modelling} of {Software}-{Intensive}
  {Systems}}, {VaMoS} '21, pages 1--9, New York, NY, USA, February 2021.
  Association for Computing Machinery.
\newblock ISBN 9781450388245.
\newblock \doi{10.1145/3442391.3442410}.
\newblock URL \url{https://dl.acm.org/doi/10.1145/3442391.3442410}.

\bibitem[Hentze et~al.(2022{\natexlab{a}})Hentze, Sundermann, Thüm, and
  Schaefer]{HST+:MODELS22}
Marc Hentze, Chico Sundermann, Thomas Thüm, and Ina Schaefer.
\newblock Quantifying the variability mismatch between problem and solution
  space.
\newblock In \emph{Proceedings of the 25th {International} {Conference} on
  {Model} {Driven} {Engineering} {Languages} and {Systems}}, {MODELS} '22,
  pages 322--333, New York, NY, USA, October 2022{\natexlab{a}}. Association
  for Computing Machinery.
\newblock ISBN 9781450394666.
\newblock \doi{10.1145/3550355.3552411}.
\newblock URL \url{https://dl.acm.org/doi/10.1145/3550355.3552411}.

\bibitem[Kuhn et~al.(2013)Kuhn, Kacker, and Lei]{KKL+:Book13}
D~Richard Kuhn, Raghu~N Kacker, and Yu~Lei.
\newblock \emph{Introduction to combinatorial testing}.
\newblock CRC press, 2013.

\bibitem[do~Carmo~Machado et~al.(2014)do~Carmo~Machado, McGregor, Cavalcanti,
  and {de Almeida}]{CM+IST14}
Ivan do~Carmo~Machado, John~D. McGregor, Yguaratã~Cerqueira Cavalcanti, and
  Eduardo~Santana {de Almeida}.
\newblock On strategies for testing software product lines: A systematic
  literature review.
\newblock \emph{Information and Software Technology}, 56\penalty0
  (10):\penalty0 1183--1199, 2014.
\newblock ISSN 0950-5849.
\newblock \doi{https://doi.org/10.1016/j.infsof.2014.04.002}.
\newblock URL
  \url{https://www.sciencedirect.com/science/article/pii/S0950584914000834}.

\bibitem[Lopez-Herrejon et~al.(2015)Lopez-Herrejon, Fischer, Ramler, and
  Egyed]{LFRE:IWCT15}
Roberto~E. Lopez-Herrejon, Stefan Fischer, Rudolf Ramler, and Aalexander Egyed.
\newblock A first systematic mapping study on combinatorial interaction testing
  for software product lines.
\newblock In \emph{2015 {IEEE} {Eighth} {International} {Conference} on
  {Software} {Testing}, {Verification} and {Validation} {Workshops} ({ICSTW})},
  pages 1--10, April 2015.
\newblock \doi{10.1109/ICSTW.2015.7107435}.
\newblock URL \url{https://ieeexplore.ieee.org/document/7107435}.

\bibitem[Chvatal(1979)]{C:MOR79}
Vasek Chvatal.
\newblock {A Greedy Heuristic for the Set-Covering Problem}.
\newblock \emph{Mathematics of operations research}, 4\penalty0 (3):\penalty0
  233--235, August 1979.
\newblock ISSN 0364-765X.
\newblock \doi{10.1287/moor.4.3.233}.
\newblock URL
  \url{https://pubsonline.informs.org/doi/abs/10.1287/moor.4.3.233}.

\bibitem[Cohen et~al.(2006)Cohen, Dwyer, and Shi]{CDS+:ISSTA06}
Myra~B. Cohen, Matthew~B. Dwyer, and Jiangfan Shi.
\newblock Coverage and adequacy in software product line testing.
\newblock In \emph{Proceedings of the {ISSTA} 2006 workshop on {Role} of
  software architecture for testing and analysis}, {ROSATEA} '06, pages 53--63,
  New York, NY, USA, July 2006. Association for Computing Machinery.
\newblock ISBN 9781595934598.
\newblock \doi{10.1145/1147249.1147257}.
\newblock URL \url{https://dl.acm.org/doi/10.1145/1147249.1147257}.

\bibitem[Perrouin et~al.(2010)Perrouin, Sen, Klein, Baudry, and
  Traon]{PS+:ICST10}
Gilles Perrouin, Sagar Sen, Jacques Klein, Benoit Baudry, and Yves~le Traon.
\newblock Automated and {Scalable} {T}-wise {Test} {Case} {Generation}
  {Strategies} for {Software} {Product} {Lines}.
\newblock In \emph{Verification and {Validation} 2010 {Third} {International}
  {Conference} on {Software} {Testing}}, pages 459--468, April 2010.
\newblock \doi{10.1109/ICST.2010.43}.
\newblock URL \url{https://ieeexplore.ieee.org/document/5477055}.
\newblock ISSN: 2159-4848.

\bibitem[Johansen et~al.(2011)Johansen, Haugen, and Fleurey]{JHF:MODELS11}
Martin~Fagereng Johansen, Øystein Haugen, and Franck Fleurey.
\newblock Properties of {Realistic} {Feature} {Models} {Make} {Combinatorial}
  {Testing} of {Product} {Lines} {Feasible}.
\newblock In Jon Whittle, Tony Clark, and Thomas Kühne, editors, \emph{Model
  {Driven} {Engineering} {Languages} and {Systems}}, pages 638--652, Berlin,
  Heidelberg, 2011. Springer.
\newblock ISBN 9783642244858.
\newblock \doi{10.1007/978-3-642-24485-8_47}.

\bibitem[Johansen et~al.(2012{\natexlab{b}})Johansen, Haugen, and
  Fleurey]{MHF:SPLC12}
Martin~Fagereng Johansen, {\O}ystein Haugen, and Franck Fleurey.
\newblock An algorithm for generating t-wise covering arrays from large feature
  models.
\newblock In \emph{Proceedings of the 16th {International} {Software} {Product}
  {Line} {Conference} - {Volume} 1}, {SPLC} '12, pages 46--55, New York, NY,
  USA, September 2012{\natexlab{b}}. Association for Computing Machinery.
\newblock ISBN 9781450310949.
\newblock \doi{10.1145/2362536.2362547}.
\newblock URL \url{https://dl.acm.org/doi/10.1145/2362536.2362547}.

\bibitem[Al-Hajjaji et~al.(2016)Al-Hajjaji, Krieter, Th\"{u}m, Lochau, and
  Saake]{AKT+:GPCE16}
Mustafa Al-Hajjaji, Sebastian Krieter, Thomas Th\"{u}m, Malte Lochau, and
  Gunter Saake.
\newblock {IncLing: Efficient Product-line Testing Using Incremental Pairwise
  Sampling}.
\newblock In \emph{Proceedings of the 2016 {ACM} {SIGPLAN} {International}
  {Conference} on {Generative} {Programming}: {Concepts} and {Experiences}},
  {GPCE} 2016, pages 144--155, New York, NY, USA, October 2016.
\newblock ISBN 9781450344463.
\newblock \doi{10.1145/2993236.2993253}.
\newblock URL \url{https://dl.acm.org/doi/10.1145/2993236.2993253}.

\bibitem[Oh et~al.(2019)Oh, Gazzillo, and Batory]{OGB+:SPLC19}
Jeho Oh, Paul Gazzillo, and Don Batory.
\newblock t-wise {Coverage} by {Uniform} {Sampling}.
\newblock In \emph{Proceedings of the 23rd {International} {Systems} and
  {Software} {Product} {Line} {Conference} - {Volume} {A}}, {SPLC} '19, pages
  84--87, New York, NY, USA, September 2019. Association for Computing
  Machinery.
\newblock ISBN 9781450371384.
\newblock \doi{10.1145/3336294.3342359}.
\newblock URL \url{https://dl.acm.org/doi/10.1145/3336294.3342359}.

\bibitem[Baranov et~al.(2020)Baranov, Legay, and Meel]{BLM+:FSE20}
Eduard Baranov, Axel Legay, and Kuldeep~S. Meel.
\newblock \emph{Baital: An Adaptive Weighted Sampling Approach for Improved
  t-Wise Coverage}, page 1114–1126.
\newblock Association for Computing Machinery, New York, NY, USA, 2020.
\newblock ISBN 9781450370431.
\newblock URL \url{https://doi.org/10.1145/3368089.3409744}.

\bibitem[Luo et~al.(2021)Luo, Sun, Qiao, Chen, Zhang, Lin, Lin, and
  Zhang]{LS+:ESEC21}
Chuan Luo, Binqi Sun, Bo~Qiao, Junjie Chen, Hongyu Zhang, Jinkun Lin, Qingwei
  Lin, and Dongmei Zhang.
\newblock {LS}-sampling: an effective local search based sampling approach for
  achieving high t-wise coverage.
\newblock In \emph{Proceedings of the 29th {ACM} {Joint} {Meeting} on
  {European} {Software} {Engineering} {Conference} and {Symposium} on the
  {Foundations} of {Software} {Engineering}}, {ESEC}/{FSE} 2021, pages
  1081--1092, New York, NY, USA, August 2021. Association for Computing
  Machinery.
\newblock ISBN 9781450385626.
\newblock \doi{10.1145/3468264.3468622}.
\newblock URL \url{https://dl.acm.org/doi/10.1145/3468264.3468622}.

\bibitem[Baranov and Legay(2022)]{BaL+:SPLC22}
Eduard Baranov and Axel Legay.
\newblock Baital: an adaptive weighted sampling platform for configurable
  systems.
\newblock In \emph{Proceedings of the 26th {ACM} {International} {Systems} and
  {Software} {Product} {Line} {Conference} - {Volume} {B}}, volume~B of
  \emph{{SPLC} '22}, pages 46--49, New York, NY, USA, September 2022.
  Association for Computing Machinery.
\newblock ISBN 9781450392068.
\newblock \doi{10.1145/3503229.3547030}.
\newblock URL \url{https://dl.acm.org/doi/10.1145/3503229.3547030}.

\bibitem[Heradio et~al.(2022)Heradio, Fernandez-Amoros, Galindo, Benavides, and
  Batory]{HF+:EMSE22}
Ruben Heradio, David Fernandez-Amoros, José~A. Galindo, David Benavides, and
  Don Batory.
\newblock Uniform and scalable sampling of highly configurable systems.
\newblock \emph{Empirical Software Engineering}, 27\penalty0 (2):\penalty0 44,
  January 2022.
\newblock ISSN 1573-7616.
\newblock \doi{10.1007/s10664-021-10102-5}.
\newblock URL \url{https://doi.org/10.1007/s10664-021-10102-5}.

\bibitem[Heß et~al.(2024)Heß, Schmidt, Ostheimer, Krieter, and
  Thüm]{HS+:VaMoS24}
Tobias Heß, Tim~Jannik Schmidt, Lukas Ostheimer, Sebastian Krieter, and Thomas
  Thüm.
\newblock {UnWise}: {High} {T}-{Wise} {Coverage} from {Uniform} {Sampling}.
\newblock In \emph{Proceedings of the 18th {International} {Working}
  {Conference} on {Variability} {Modelling} of {Software}-{Intensive}
  {Systems}}, {VaMoS} '24, pages 37--45, New York, NY, USA, February 2024.
  Association for Computing Machinery.
\newblock ISBN 9798400708770.
\newblock \doi{10.1145/3634713.3634716}.
\newblock URL \url{https://dl.acm.org/doi/10.1145/3634713.3634716}.

\bibitem[Hentze et~al.(2022{\natexlab{b}})Hentze, Pett, Sundermann, Krieter,
  Thüm, and Schaefer]{HPS+:GPCE22}
Marc Hentze, Tobias Pett, Chico Sundermann, Sebastian Krieter, Thomas Thüm,
  and Ina Schaefer.
\newblock Generic {Solution}-{Space} {Sampling} for {Multi}-domain {Product}
  {Lines}.
\newblock In \emph{Proceedings of the 21st {ACM} {SIGPLAN} {International}
  {Conference} on {Generative} {Programming}: {Concepts} and {Experiences}},
  {GPCE} 2022, pages 135--147, New York, NY, USA, December 2022{\natexlab{b}}.
  Association for Computing Machinery.
\newblock ISBN 9781450399203.
\newblock \doi{10.1145/3564719.3568695}.
\newblock URL \url{https://dl.acm.org/doi/10.1145/3564719.3568695}.

\end{thebibliography}





% \end{thebibliography}

\end{document}